\documentclass{aa}  
\usepackage{graphicx}
\usepackage{color}
\definecolor{esoblue}{RGB}{0,119,190}
\definecolor{lightblue}{RGB}{231,243,255}
\definecolor{darkblue}{RGB}{0,1,70}
\definecolor{dblue}{RGB}{51,51,153}
\definecolor{lblue}{RGB}{0,102,255}
\definecolor{darkgreen}{RGB}{64,159,64}
\definecolor{lightgreen}{RGB}{207,231,207}
\definecolor{lightgreen1}{RGB}{152,255,97}
\definecolor{lightyellow}{RGB}{255,255,78}
\definecolor{darkred}{RGB}{191,0,0}
\definecolor{lightred}{RGB}{249,229,229}
\definecolor{lightred1}{RGB}{255,117,117}
\definecolor{orange}{RGB}{247,180,31}
\definecolor{pink}{RGB}{255,0,240}
\usepackage[%
colorlinks,%
linkcolor=dblue,
citecolor=dblue%
]{hyperref}

\usepackage{txfonts}
\begin{document}

   \title{CRIRES$^+$ on sky at the ESO Very Large Telescope}
   \subtitle{Observing the Universe at infrared wavelengths and high spectral resolution}

   \author{R. J. Dorn\inst{1}, P. Bristow \inst{1}, J.V. Smoker \inst{2}, F. Rodler\inst{2}, A. Lavail\inst{3}, M. Accardo \inst{1}, M. van den Ancker\inst{1}, D. Baade\inst{1}, A. Baruffolo\inst{4}, B. Courtney-Barrer\inst{2}, L. Blanco\inst{2}, A. Brucalassi\inst{5}, C. Cumani \inst{1}, R. Follert\inst{6}, A. Haimerl\inst{1}, A. Hatzes\inst{6}, M. Haug \inst{1}, U. Heiter\inst{3}, R. Hinterschuster\inst{1}, N. Hubin\inst{1}, D. J. Ives \inst{1}, Y. Jung \inst{1}, M. Jones\inst{2}, J-P. Kirchbauer \inst{1}, B. Klein  \inst{1}, O. Kochukhov\inst{3}, H. H. Korhonen\inst{2}, J. Köhler\inst{6}, J-L. Lizon \inst{1}, C. Moins \inst{1}, I. Molina-Conde \inst{1}, T. Marquart\inst{3}, M. Neeser \inst{1}, E. Oliva\inst{5}, L. Pallanca\inst{2}, L. Pasquini \inst{1}, J. Paufique \inst{1}, N. Piskunov\inst{3}, A. Reiners\inst{7}, D. Schneller \inst{1}, R. Schmutzer\inst{2}, U. Seemann \inst{1}, D. Slumstrup\inst{2}, A. Smette\inst{2}, J. Stegmeier \inst{1}, E. Stempels\inst{3}, S. Tordo \inst{1}, E. Valenti \inst{1}, J. J. Valenzuela\inst{2}, J. Vernet\inst{1}, J. Vinther \inst{1}, and A. Wehrhahn\inst{3}}

   \institute{European Southern Observatory, Karl-Schwarzschild Str. 2, 85748 Garching b. München, Germany\\
              \email{rdorn@eso.org}
        \and European Southern Observatory, Alonso de Cordova 3107, Vitacura, Santiago, Chile\
        \and Department of Physics and Astronomy, Uppsala University, Box 516, 751 20 Uppsala, Sweden\
        \and INAF Osservatorio Astronomico di Padova, Vicolo Osservatorio 5, 35122 Padua, Italy\
        \and INAF Arcetri Osservatorio, Largo E. Fermi 5, 50125 Firenze, Italy\
        \and Thüringer Landessternwarte Tautenburg, Sternwarte 5, 07778 Tautenburg, Germany\
        \and Institut für Astrophysik, Universität Göttingen, Friedrich-Hund-Platz 1, 37077 Göttingen, Germany\\
}
   \date{Received October 13, 2022; accepted December 26, 2022}

\titlerunning {CRIRES$^+$ on sky at the ESO Very Large Telescope}
\authorrunning {R. J. Dorn et al.}


  \abstract
   {The CRyogenic InfraRed Echelle Spectrograph (CRIRES) Upgrade project CRIRES$^+$ extended the capabilities of CRIRES.
   It transformed this VLT instrument into a cross-dispersed spectrograph to increase the wavelength 
   range that is covered simultaneously by up to a factor of ten. In addition, a new detector focal plane 
   array of three Hawaii 2RG detectors with a 5.3 $\mu$m cutoff wavelength replaced the existing 
   detectors. Amongst many other improvements, a new spectropolarimetric unit was added and the 
   calibration system has been enhanced. The instrument was installed at the VLT on Unit Telescope 3 
   at the beginning of 2020 and successfully commissioned and verified for science operations during 2021, 
   partly remotely from Europe due to the COVID-19 pandemic. The instrument was subsequently offered 
   to the community from October 2021 onwards. This article describes the performance and capabilities 
   of the upgraded instrument and presents on sky results.}


   \maketitle
    \section{Introduction}
    High-resolution infrared (IR) spectroscopy plays an important role in astrophysics from the search for exoplanets to cosmology. The majority of currently existing IR spectrographs are limited by their small simultaneous wavelength coverage. The scientific community has recognized the need for large wavelength range, high-resolution IR spectrographs, and several are currently either in the design, integration phase, or at the telescope. Examples are ISHELL \citep{2012SPIE.8446E..2CR} at the NASA Infrared Telescope Facility, SPIRou \citep{2020MNRAS.498.5684D} for the Canada France Hawaii Telescope (CFHT), GIANO at the TNG telescope \citep{2014SPIE.9147E..1EO}, CARMENES for the Calar Alto Observatory \citep{2012SPIE.8446E..0RQ} and the Near Infra-Red Planet Searcher (NIRPS) at the ESO 3.6-metre telescope in La Silla \citep{2017SPIE10400E..18W}.
    
    The Adaptive Optics (AO) assisted CRyogenic InfraRed Echelle Spectrograph (CRIRES) instrument, previously installed at the Very Large Telescope (VLT), was an IR (0.92 - 5.3 $\mu$m) high-resolution spectrograph which was in operation from 2006 to 2014. CRIRES was a unique instrument, accessing a parameter space (wavelength range and spectral resolving power), which up to then was largely uncharted, as described by \citet{2004SPIE.5492.1218K} and for the AO system presented in \citet{2004SPIE.5490..216P}. In its setup, it consisted of a single-order spectrograph providing long-slit (40 arcseconds) spectroscopy with a resolving power up to $R$=100,000. However the setup was limited to a narrow, single-exposure, spectral range of about 1/70 of the central wavelength, resulting in low observing efficiency for many modern scientific programmes requiring a broad spectral coverage. By introducing cross-dispersing elements and larger detectors, the simultaneous wavelength range was increased up to a factor of ten with respect to the old configuration, while the total operational wavelength range was preserved. 
    
   \begin{figure*}
   \centering
   \includegraphics[width=0.9\textwidth]{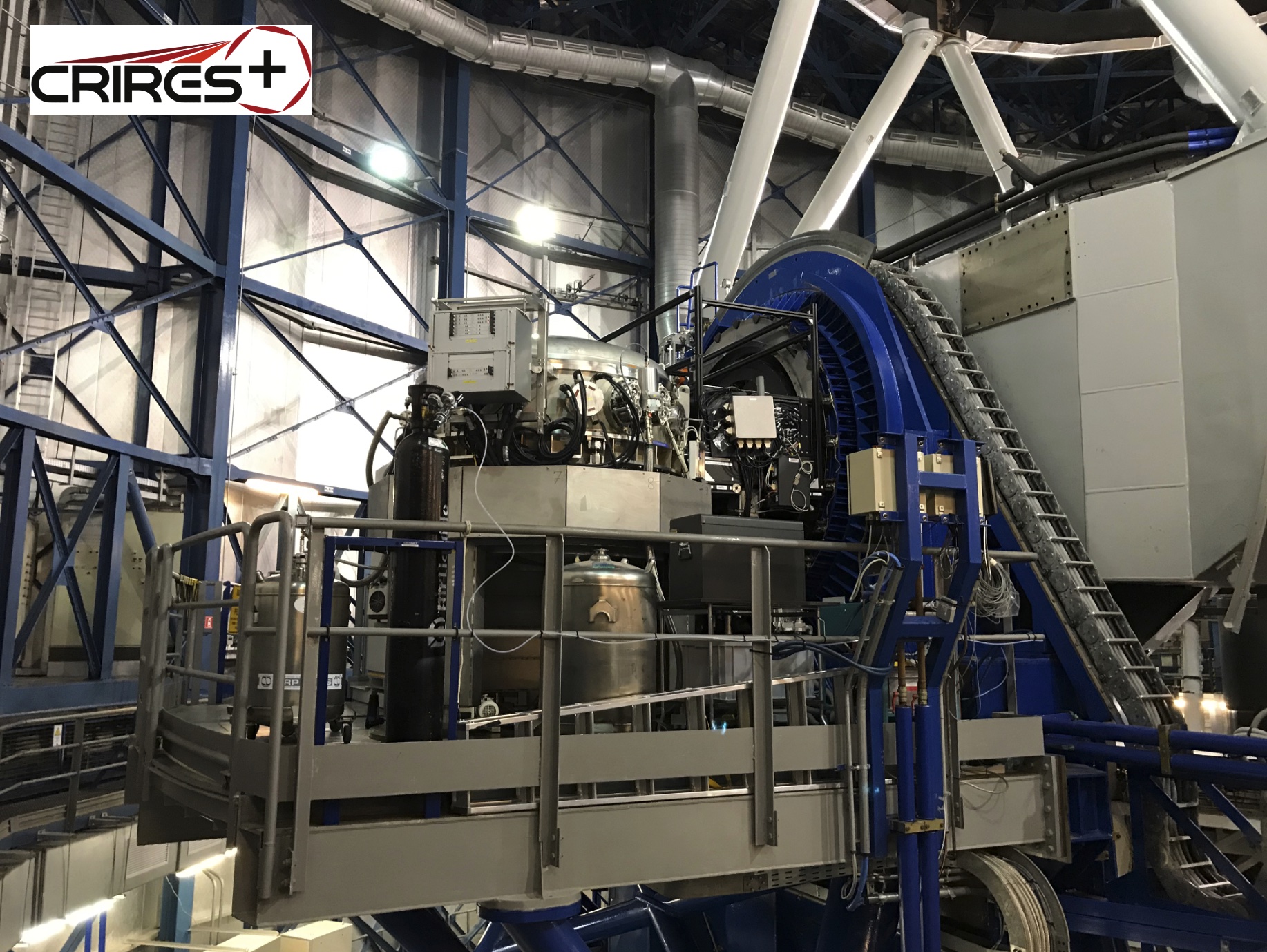}
   \caption{CRIRES$^+$ instrument in February 2020 installed on Nasmyth focus B of Unit Telescope 3 at the ESO VLT.}
    \label{fig:instrument}
    \end{figure*}
   
\section{Science drivers}

A select number of science drivers for the CRIRES Upgrade project (CRIRES$^+$) are listed below. Others such as observations of brown dwarfs, the search for Hydrogen Bracket and Pfund series in supergiant stars and observations of solar system bodies are described in \cite{2022Msngr.187...17L}.

\subsection{A search for super-Earths in the habitable zone of low-mass stars}
    At the time of writing, approximately 20 percent of exoplanets\footnote{https://exoplanets.nasa.gov/exoplanet-discoveries/} 
    have been discovered through radial velocity (RV) measurements. However, only 5\% of the planets detected so far orbit stars with stellar masses of less than about 0.5~$M_{\rm sun}$. Thus, we lack key knowledge about the process of planet formation around the most numerous stars in our galaxy – M dwarfs. M dwarfs comprise some $\sim$80 percent of stars in the Milky Way (\cite{Winters2019} and references therein) and are a prime objective of current and future near-IR spectrometers. Low-mass stars are especially interesting because these objects are cool and the habitable zones are quite close to the star. The reflex motion of an M star (0.15 $M_{\rm sun}$) with a 1 $M_{\rm Earth}$ planet in its habitable zone is about 1~m~s$^{-1}$. Since M dwarfs and brown dwarfs have low effective temperatures, radiating most of their energy in the IR from 1.0 to 2.5~$\mu$m  \citep{2016PhR...663....1S}, a high-resolution IR spectrograph is therefore ideal for searching for low-mass planets around these objects. A new gas absorption cell as described in section 3.3 provides a stable wavelength reference which, combined with the increase in wavelength coverage by about a factor of ten, should now result in an attainable RV precision for CRIRES$^+$ of 3~m~s$^{-1}$ or better. This could enable the detection of super-Earth-mass planets in the habitable zone of an M-dwarf star in the solar neighbourhood.
    
\subsection{Atmospheric characterization of transiting planets}
    In-transit spectroscopy of exoplanets currently provides us with one of the most powerful means of studying exoplanetary atmospheres \citep{2009ApJ...698..519K}. Transiting planets are almost always worlds that happen to be close to their host stars; they are hot and radiate most of their light in the IR. Furthermore, the IR is a spectral region where molecular gases, such as CO, CO$_2$, NH$_3$, CH$_4$, H$_2$O, among others, produce strong spectral features. Observations of transiting exoplanets with CRIRES$^+$ are presented in \cite{2022Msngr.187...17L} and \cite{2022AJ....164...79H}. In addition to the 1.0 to 2.5 $\mu$m domain (where several of these molecules have strong transitions and several instruments operate), CRIRES$^+$ is one of the very few instruments that covers the 3.0 to 5.2 $\mu$m domain at high-spectral resolution. This important wavelength region allows for the observation of multiple gases in exoplanetary atmospheres simultaneously.
    
\subsection{Origin and evolution of stellar magnetic fields}

    Magnetic fields play a fundamental role in the life of all stars: they govern the emergence of stars from proto-stellar clouds, control the infall of gas onto the surfaces of young stars, and aid the formation of planetary systems. Thanks to modern high-resolution spectrographs and spectropolarimeters, stellar magnetic fields are being routinely observed and characterized across the H-R diagram from cool to hot stars \citep{2009ARA&A..47..333D, 2012LRSP....9....1R, 2021A&ARv..29....1K}. Current and upcoming high-resolution spectropolarimeters operating in the near-infrared regime -- such as SPIRou at the CFHT -- or SPIP at the T\'elescope Bernard Lyot are installed at respectively 4-m and 2-m class telescopes. CRIRES$^+$, which benefits from the 8-m large collecting area of the VLT, will allow the study of the magnetism of fainter objects, as well as poorly studied Class 0 and Class I objects, which are still embedded and thus very challenging to observe in the visible. 
    
    CRIRES$^+$ makes it possible to measure with greater accuracy magnetic fields in cool stars and ultra-cool dwarfs. The Zeeman splitting of a spectral line is proportional to $\lambda^2$, so there is a huge leverage in going to the IR. For cool objects most of the flux is in the IR so there is also a gain due to the increased signal-to-noise ratio. In order to disentangle Zeeman broadening from other broadening effects one must compare the broadening of Zeeman sensitive lines to magnetically insensitive lines. The large wavelength coverage of CRIRES$^+$ includes many more lines of different magnetic sensitivities needed for an accurate determination of the field strength. CRIRES$^+$ operates in the infrared and hence enables observations to be performed in dusty regions where optical light is heavily affected by extinction. In addition, the capability of CRIRES$^+$ to take circular and linear polarized spectra can support these measurements.

\subsection{Protoplanetary discs}

The formation of giant planets is thought to occur in the inner regions of protoplanetary discs \citep{Armitage2011}. The original CRIRES was used to investigate CO line in the innermost $\sim$10 au of such discs by employing the technique of spectroastrometry \citep{Pontoppidan2011,Brown2013}. Several other organic molecules in T-Tauri stars were also detected using the original instrument \citep{2012ApJ...747...92M}. Follow up observations using CRIRES$^+$ \citep{2022Msngr.187...17L} were executed in order to study the composition, kinematics and structure of the binary system S CRa. The extended wavelength coverage in the L and M bands allowed the detection of H$_{2}$O, OH, HCN, and CO and a number of atomic hydrogen lines. These and similar observations will be used to study the envelopes and possible jets and outflow velocities of protoplanetary discs and determine their evolutionary status and ability to form planets. Finally, CRIRES$^+$ observations of the Herbig Ae star HD\,36917 presented in \cite{2022Msngr.187...17L} show CO emission around 2.3 $\mu$m and compare it with a Keplerian disc model.


    In summary, the scientific cases for high-resolution infrared spectroscopy were strong and subsequently the ESO Scientific Technical Committee recommended the upgrade of CRIRES after the completion of the detailed Phase A study in April 2013. The ESO Council at its 128$^{\rm th}$ meeting in June 2013 then endorsed this proposal. Finally the instrument was installed at the VLT  at the beginning of 2020 (see Fig. \ref{fig:instrument}) although the commissioning activities had to be put on hold due to the temporary closure of the Paranal Observatory due to COVID-19. The commissioning runs were then executed remotely using the Garching Remote Access facility (G-RAF) to the control room in the VLT Interferometer complex. It was the first time that such a remote commissioning took place at ESO. Whilst the CRIRES$^+$ commissioning runs contributed considerably to the development of practices for remote instrument operation, characterization and commissioning at ESO, this way of working caused a significant overhead for the project. After commissioning, the CRIRES$^+$ instrument has been in regular operations since 1 October 2021.

\section{CRIRES$^+$ instrument description}

    Many astrophysical applications benefit significantly from the increase in wavelength coverage introduced by turning CRIRES into a cross-dispersed echelle spectrograph. The CRIRES upgrade project improved the instrument by either refurbishing or replacing numerous subsystems. Furthermore, it provides additional observing modes compared to the original instrument. The project identified the following upgrades as significantly impacting the scientific capabilities of CRIRES. The main drivers for this upgrade and the initial designs are described in articles by \citet{2014Msngr.156....7D, 2014SPIE.9147E..19F}. A summary of the new and main instrument parameters is given in Table \ref{tab1}.

   \begin{figure*}
   \centering
   \includegraphics[width=2\columnwidth]{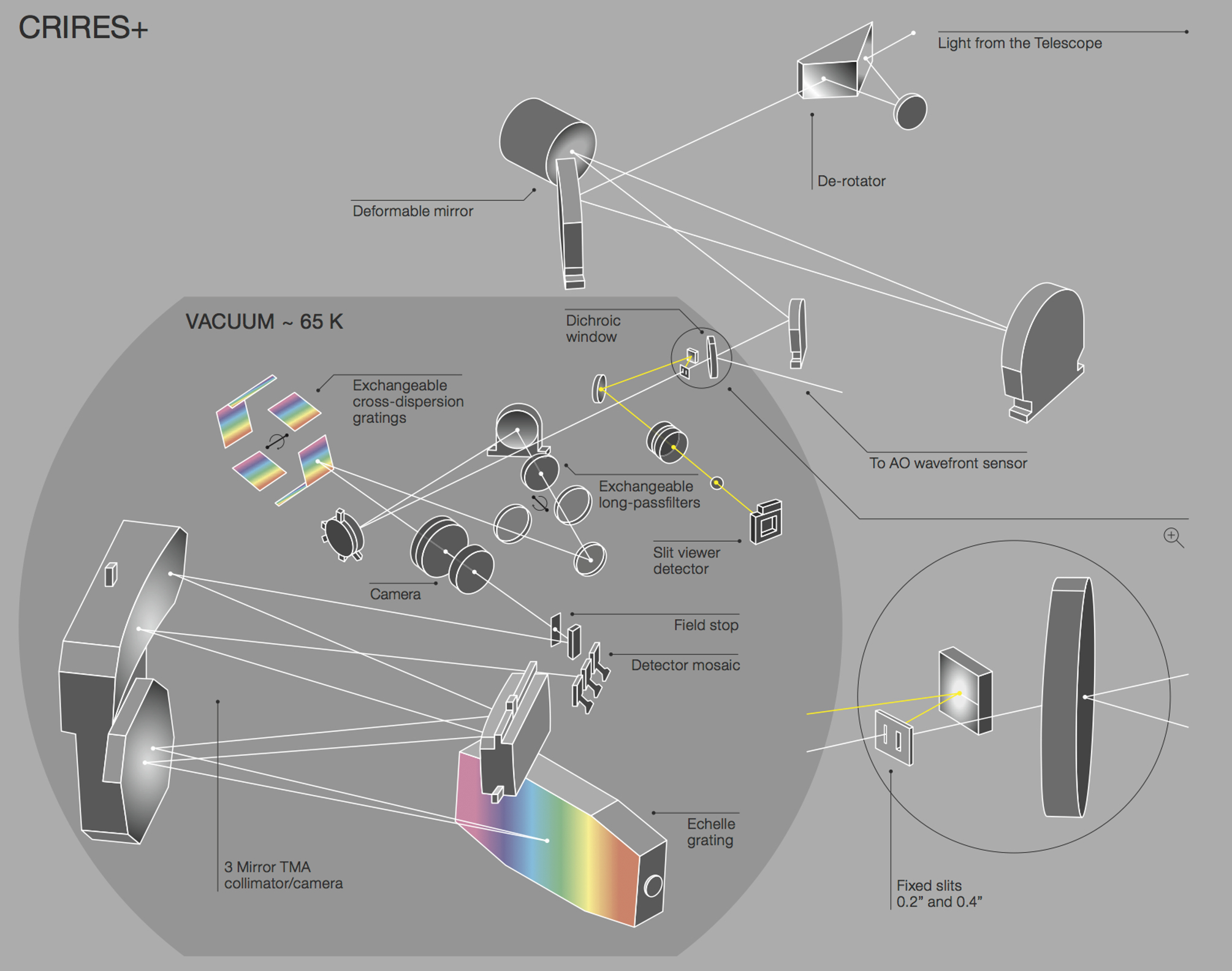}
   \caption{CRIRES$^+$ instrument optical design including the new cross-dispersion pre-optics.}
   \label{fig:design}
   \end{figure*}
   
\subsection{Transform CRIRES into a cross dispersed echelle spectrograph}

    To cover the additional orders, the spatial extent of the main slit was reduced from 40 to 10 arcseconds, providing a balanced compromise between cross-dispersion implementation and catering for the current CRIRES long slit usage. The cross-dispersion of the spectrum is performed by the use of reflection gratings. The optical layout of the new design is shown in Fig. ~\ref{fig:design}. There are six gratings mounted on a cryogenic wheel. Each of them is optimized for operation in a single wavelength band (Y, J, H, K, L and M). Another wheel carries the order-sorting filters to eliminate contamination by second- and higher- order spectra of the gratings. The re-imaging of the slit is then performed by a fixed-lens camera, designed for the full wavelength coverage and used for all observing modes. In this configuration, the observing modes require two exposures to cover the full Y-band with different echelle angles, three for the J-band, four for the H and K-bands, seven for the L-band and nine exposures are needed to cover the M-band. In total, 29 settings are needed to cover YJHKLM compared with around 280 settings for the original CRIRES. A new optical design for the slit viewer complemented the new pre-optics unit. Overall, the new design was meant to maintain the current throughput of CRIRES (with the goal of increasing it) and the spectral resolving power of $\sim$50,000 and $\sim$100,000, as before. The slit length (10 arcseconds) does not limit observations of moderately extended sources and allows nodding for precise background subtraction. 

\subsection{Optomechanical design of the new pre optics unit }

\begin{figure}
   \centering
   \includegraphics[width=1\columnwidth]{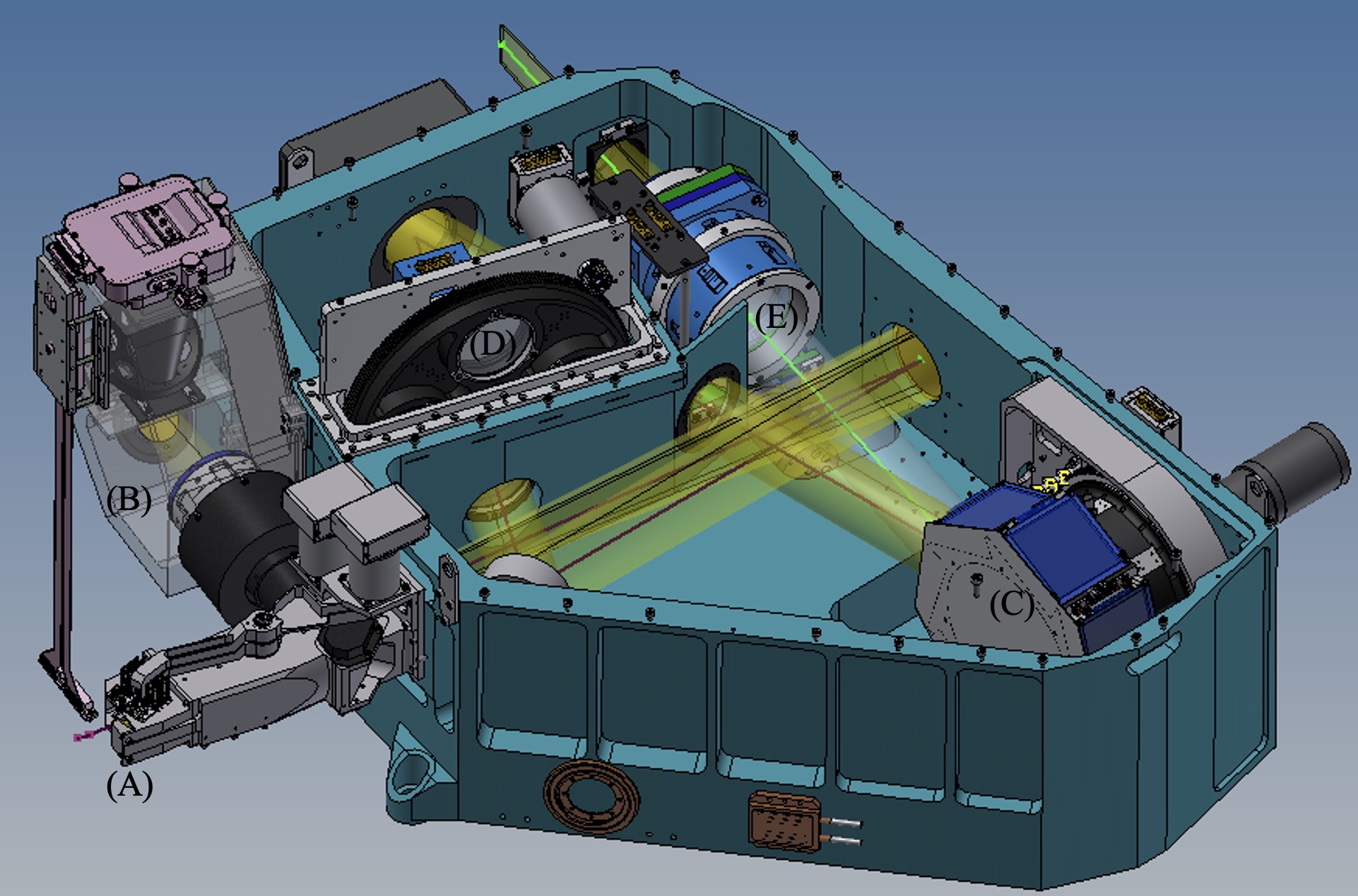}
      \caption{Top view of the new CRIRES$^+$ fore-optics assembly with heat exchangers, jitter mirror unit, SV camera subsystem, slit and decker subsystem, CDU subsystem, camera unit and one of the fixation points. 
              }
         \label{fig:foreoptics}
   \end{figure}

    The fore-optics of the original CRIRES was completely renewed as shown in Fig.~\ref{fig:foreoptics}. It was replaced by an off-axis parabola, which creates a collimated beam with a diameter of 50 mm, followed by two flat mirrors with distances and angle adjusted to match the new fore-optics with the already existing three-mirror anastigmat (TMA) relay optics and the echelle grating remained from the original CRIRES. A new entrance slit (A) was implemented at a location inside just before the entrance window. The entrance window is a dichroic reflecting visible light for the AO system. Light not passing through the entrance slit is reflected to the upgraded slit viewer system (B). The slit unit comprises a movable mask with two slits: 0.2” (resolving power $\sim$100,000) and a 0.4” slit (resolving power $\sim$50,000). The mask can also be positioned so that neither slit is in the optical path and the spectrograph is closed to light from the telescope. The reproducibility and stability are significantly enhanced compared to the original slit mechanism. In addition, the CRIRES$^+$ entrance slit mechanism includes a decker for polarimetric observations allowing for the left and right-hand polarized beams at two nodding positions. 

  \begin{table}
      \caption[]{Summary of new and main instrument parameters}
      \label{tab1}
\small
\begin{tabular}{l l} 
 \hline
Spectral resolution& $\sim$50,000 and $\sim$100,000\\ 
  \hline
Wavelength coverage& 0.95 - 5.3 $\mu$m (YJHKLM bands)\\ 
 \hline
RV precision&	goal: 3~m~s$^{-1}$\\
 \hline
Slit length & 	10 arcseconds\\
\hline
Slit width	& 0.2 and 0.4 arcseconds\\
\hline
Polarimetry& linear + circular (YJHK bands)\\
\hline
Adaptive Optics&	60 actuator curvature sensing\\
\hline
Cross-disperser&	6 gratings\\
\hline
Focal plane& 6144x2048 pixels (Mosaic of three H2RGs)\\
\hline
\end{tabular}
   \end{table}
   
    As for the original CRIRES, the aim of the slit viewer is to perform the acquisition of the observed target and to measure the offset with respect to the tracking position on the slit. However, the slit viewer subsystem is substantially modified. The CRIRES$^+$ slit viewer subsystem is composed of two folding mirrors, a camera to image the entrance slit on a detector and a filter wheel to select the filter for guiding. Relative to the original CRIRES, the Aladdin detector is replaced with an engineering grade H2RG which significantly enhanced the SV camera performance.  
   
\begin{figure}
   \centering
   \includegraphics[width=1\columnwidth]{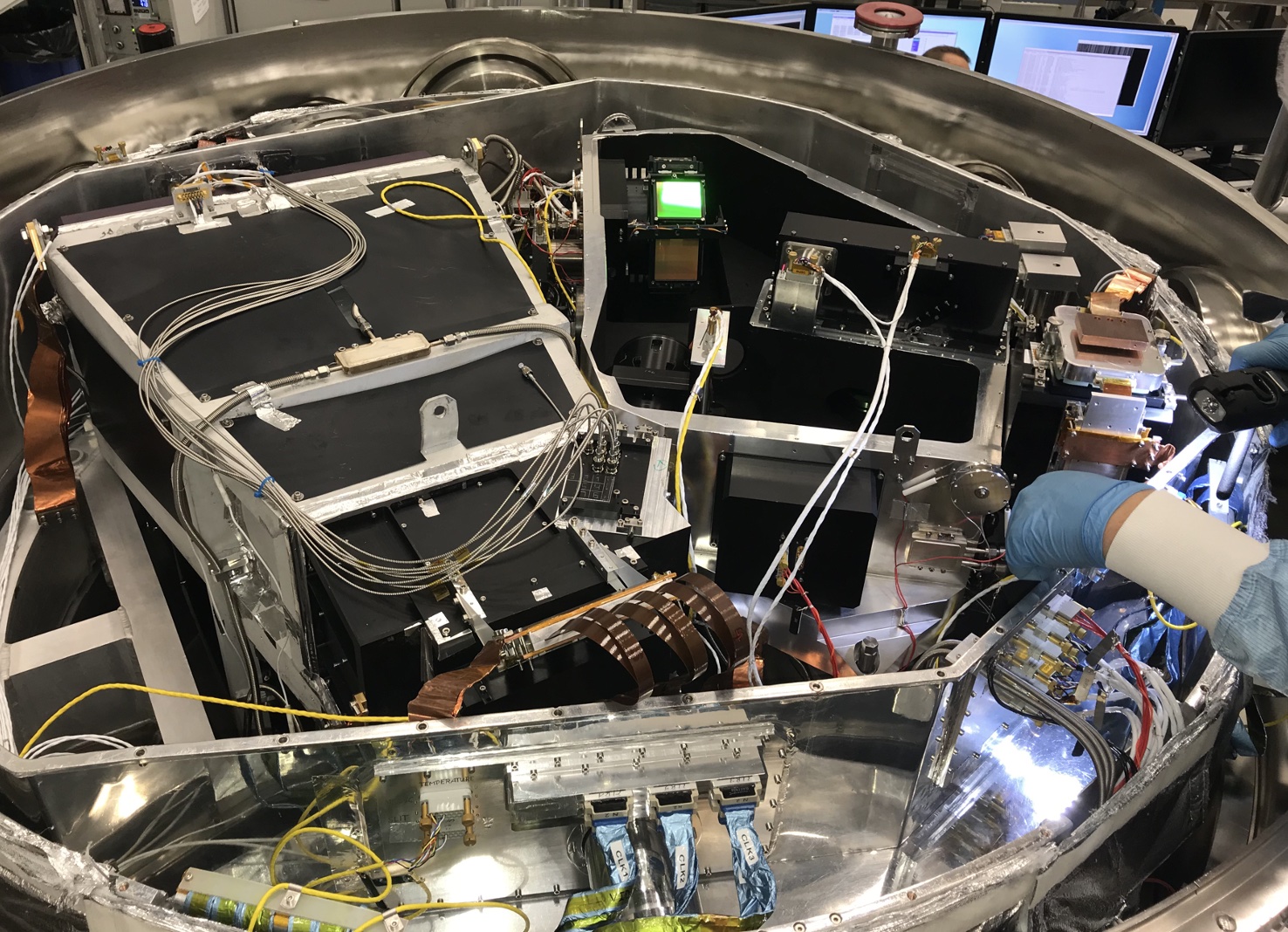}
      \caption{Actual implementation of the new cross dispersion optics and detectors in the instrument
              }
         \label{fig:implem}
   \end{figure}

    The old pre-disperser prism is replaced by a cross-disperser subsystem. As can be seen in Fig.~\ref{fig:implem}, the beam from the f/15 focus at the new entrance slit is collimated by a parabolic mirror and arrives at the cross-disperser wheel (C) via two flat mirrors and a long pass filter to block the 2$^{\rm nd}$ and higher orders of the cross-disperser gratings. The jitter mirror has two piezo actuators that allow the echellogram to be translated at sub-pixel accuracy on the detectors. The order-sorting filter (D) can be selected from one of three filters (or an open position) on a wheel to be appropriate to the chosen cross-disperser grating (in pairs YJ, HK, LM). The cross-disperser wheel contains six reflection gratings, one for each of the bands Y, J, H, K, L and M. To measure the grating efficiency and quality, the National Metrology Institute of Germany (Physikalisch-Technische Bundesanstalt; PTB) Berlin was contracted to characterize all foreseen gratings for CRIRES$^+$. The results on characterizing the cross dispersion reflection gratings of CRIRES$^+$ are described in more detail by \citet{2016SPIE.9912E..2BF}. 

\begin{figure}
   \centering
   \includegraphics[width=1\columnwidth]{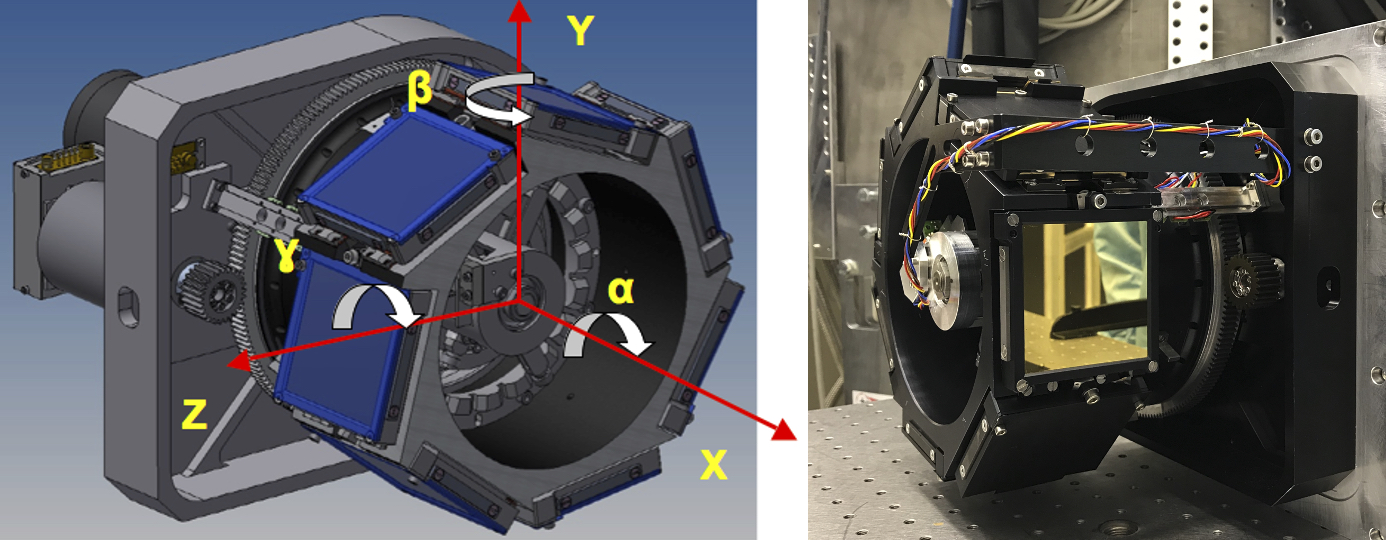}
      \caption{Grating wheel design with locking mechanism (abandoned and not used in the instrument) and build prototype to verify repeatability and stability.
              }
         \label{fig:gratingwheel}
   \end{figure}

    A locking mechanism was supposed to ensure accurate repeatability of the cross-disperser wheel. During the design phase the grating wheel unit as shown in Fig.~\ref{fig:gratingwheel} was identified as one critical function of the instrument. Its operation and more specifically its positioning accuracy was very demanding. Therefore, it was decided to build a prototype of this function in order to assess its real feasibility and demonstrate early enough the ability to meet the positioning accuracy and repeatability. Unfortunately, during the system testing phase in the laboratory of the instrument the locking mechanism showed various problems and varying positioning accuracy when moving and locking. It was decided to abandon the locking function of the wheel and the accurate positioning was then achieved by the implementation of a novel metrology system, commanding the grating wheel positioning, the piezo mirror function as well as the echelle grating function. When metrology is applied, the spectral format is reproducible in the dispersion direction to better than 0.1 pixels (in wavelength units: $\sim$0.0003 nm in Y-band; $\sim$0.0014 nm in M-band) and in cross dispersion to better than 0.5 pixels on the focal plane detectors.

    Following the cross-disperser grating, an achromatic camera (E) working at a fixed focal length brings the collimated beam to an f/8 focus at the field stop. In order to avoid time consuming thermal cycling during the Assembly, Integration and Testing (AIT) phase the camera is mounted on a small and simple focusing stage. This focusing functionality is only intended for maintenance, not for regular operations. Finally a piezo-driven flat mirror allows fine positioning of the spectral format on the detector array. The echelle grating subsystem is unchanged relative to the original CRIRES.  More details on the optical and opto-mechanical designs can be found in articles by \citet{2014SPIE.9147E..7SL} and \citet{2014SPIE.9147E..7RO}, respectively. 

\subsection{Gas cells and Uranium-Neon lamp for a new level of wavelength calibration}
\begin{figure}
   \centering
   \includegraphics[width=1\columnwidth]{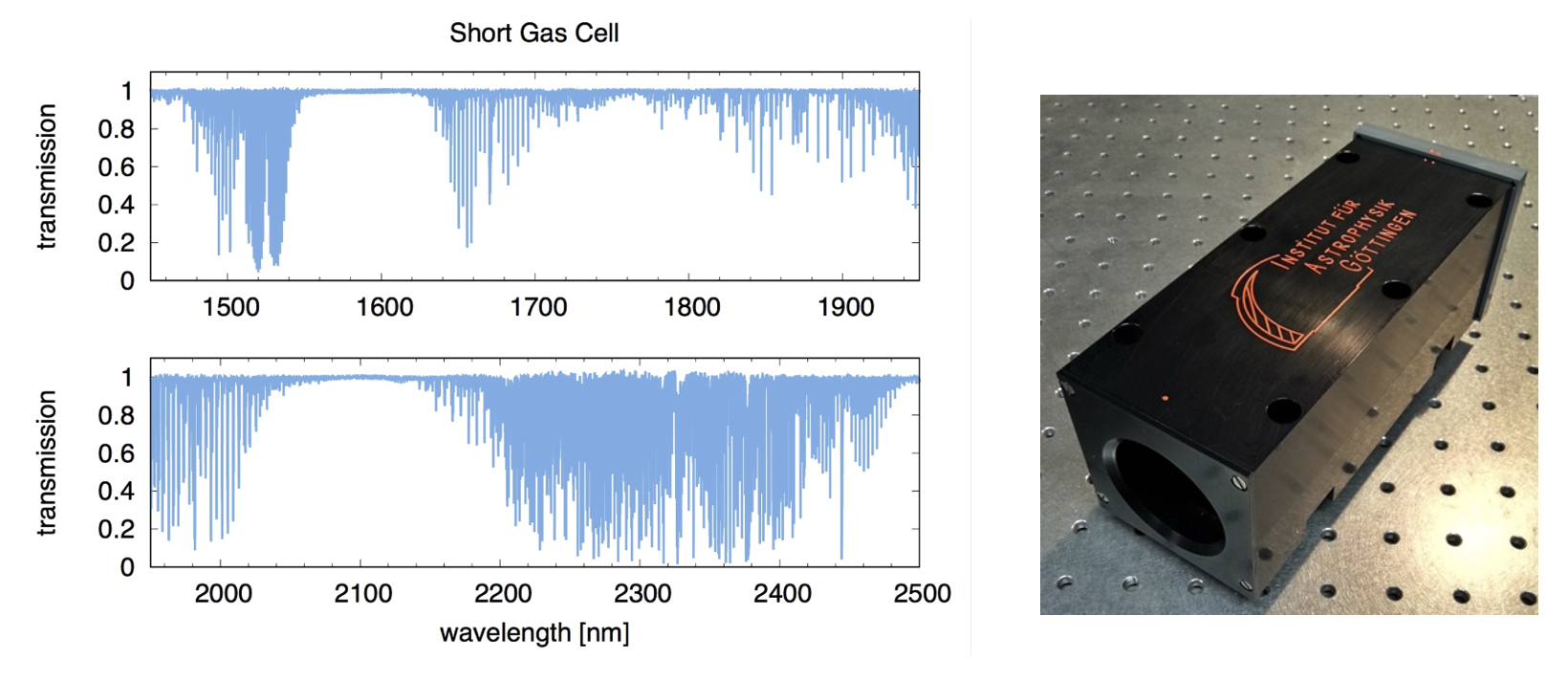}
      \caption{Absorption line spectrum of the new short gas cell (SGC) in the H and K bands on the left and the gas cell on the right. 
              }
         \label{fig:gascell}
   \end{figure}

    The CRIRES$^+$ science cases demand specialized, highly accurate wavelength calibration techniques. Therefore, another part of the upgrade was concerned with the installation of novel IR absorption gas cells with multi-species gas fillings (NH$_3$, $^{13}$CH$_4$, C$_2$H$_2$). These gases provide a set of densely distributed absorption lines imprinted on the stellar spectra in the H- and K-bands (see Fig.~\ref{fig:gascell}). In addition, the existing Thorium-Argon hollow cathode lamp was replaced with a similar Uranium-Neon (UNe) lamp that produces a richer wavelength calibration spectrum. 

\subsection{Fabry-Perot interferometer}

\begin{figure}
   \centering
   \includegraphics[width=1\columnwidth]{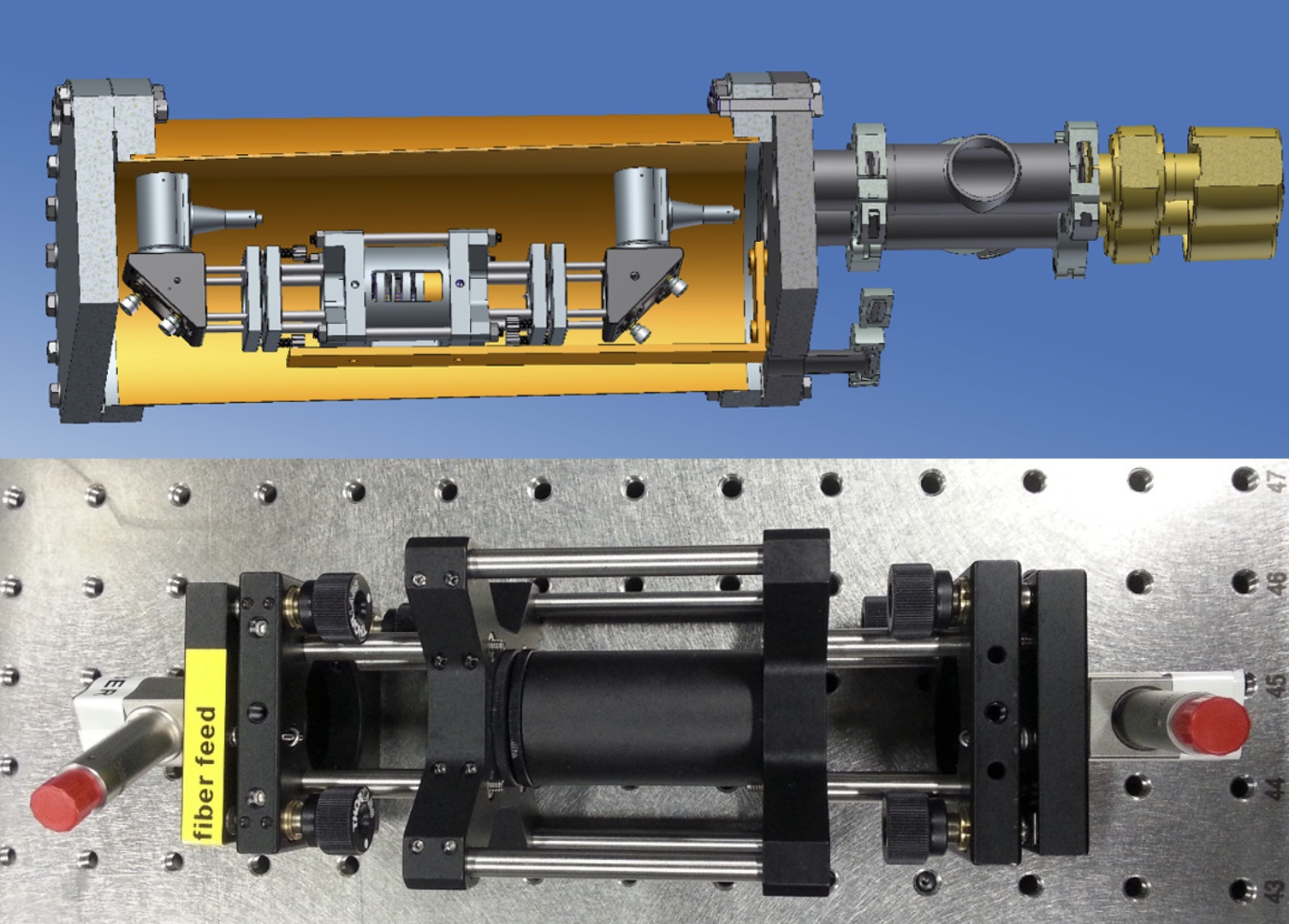}
      \caption{Fabry-Perot interferometer model (top) and demonstrator (bottom) for CRIRES$^+$.
              }
         \label{fig:etalon}
   \end{figure}
   
    An etalon for CRIRES$^+$ is an addition that was recommended during the preliminary design review. Such an additional wavelength calibration device covering Y, J, H and K-bands mitigates shortcomings of other devices such as the hollow cathode lamp. A Fabry-Perot etalon (or Fabry-Perot interferometer, FPI) as shown in Fig.~\ref{fig:etalon} can be used to create a periodic signal in frequency space by means of interference. Each of these fringes serves as a reference marker to tackle the wavelength calibration. For this purpose, a continuum light source with a feature free, flat broadband spectrum is coupled to a Fabry-Perot cavity, where interference is produced. The choice of cavity length and the properties of the cavity windows/mirrors (finesse $F$) determine the peak separation (free spectral range, FSR) and the line strength (sharpness, contrast). The FSR and contrast can thus be tuned and optimized to match the spectrograph's resolving power, sampling, and wavelength range. The major advantages are comb-like, equidistant reference lines over the design range with high homogeneity, equally strong spectral features, thus homogeneous line contrast, broadband coverage with no gaps and a high line density. The absolute wavelength calibration is anchored to lines provided by the Uranium-Neon lamp. 

    The FPI subsystem comprises a sealed vacuum vessel, standard ESO vacuum pump and a halogen light source. Sub-atmospheric pressure is achieved by daily pumping (duration around 30 mins) to 10$^{-3}$ mbar, this pumping process is independent of the main spectrograph cryo-vacuum subsystem. An interlock valve closes the FPI chamber in case of pump failure. All three components are secured on a bench in the base of the warm structure as illustrated in Fig.~\ref{fig:warmpart}. As can be seen, the base is attached to the warm structure, not directly to the Nasmyth platform. The FPI feeds a fibre which delivers the FPI spectrum to the integrating sphere. The calibration system is described in more detail by \citet{2014SPIE.9147E..5GS}.  
    
    The stability of the FPI has been characterized in the laboratory against a stabilized frequency standard (iodine). The intrinsic stability of the FPI against this reference under controlled conditions over $>$30 hours is measured at 1.3 m~s$^{-1}$ RMS, over 60 hours at 3.3 m~s$^{-1}$ RMS. We note that the stability required for zero-pointing the FPI, that is assigning to the fringes an absolute wavelength, is only relevant on short timescales of minutes. Exposures of the FPI and an absolute wavelength standard are always adjacent for wavelength calibration data products, and hence the FPI stability is of relevance only over the length of this calibration sequence. The intrinsic drift that the FPI experiences as a result of residual temperature variations during such a sequence is governed by the induced ambient temperature gradient on the Nasmyth platform over this period, and is of the order of 1 m~s$^{-1}$ typically. The drift over a typical observing night (8 hours) is within  \textpm 5 m~s$^{-1}$, over 24 hours within \textpm 10 m~s$^{-1}$, under nominal operating conditions.

\subsection{New state-of-the-art detectors}

\begin{figure}
   \centering
   \includegraphics[width=1\columnwidth]{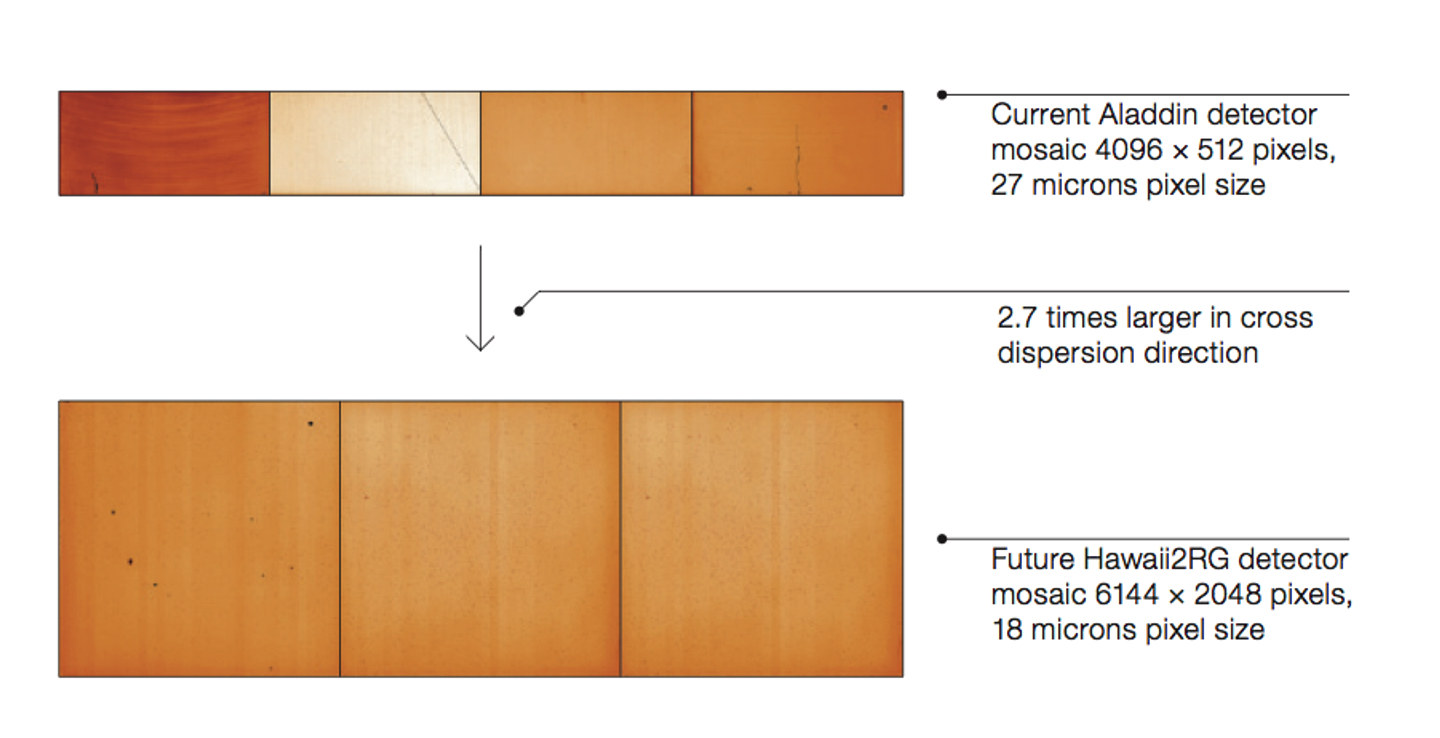}
      \caption{Original CRIRES detector mosaic focal plane array area (top) compared to the new detectors (bottom) with an increase of a factor of 2.7 in cross dispersion direction.
              }
         \label{fig:detsize}
   \end{figure}

\begin{figure}
   \centering
   \includegraphics[width=1\columnwidth]{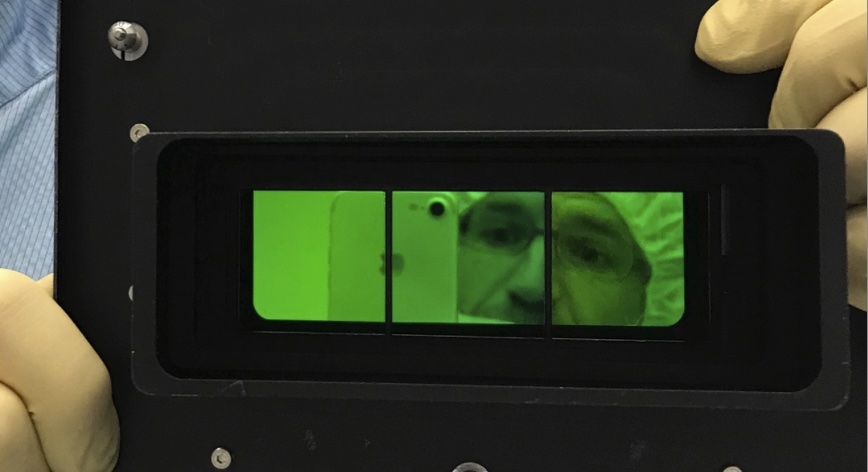}
      \caption{Three CRIRES$^+$ H2RG detectors arranged as a linear mosaic in the CRIRES$^+$ specific detector mount including the light baffle. 
              }
         \label{fig:det}
   \end{figure}
   
    Another major part of the upgrade project is to increase the coverage of the focal plane by introducing a set of new detectors. The origional CRIRES scientific detector system used four Raytheon 1024 x 1024 pixels InSb Aladdin arrays \citep{2004SPIE.5499..510D}. Another Aladdin detector was used for the slit-viewer camera. Owing to the cross dispersion, a larger field was required in CRIRES$^+$ to cover up to ten orders per band with a slit length of 10 arcseconds. Fig.~\ref{fig:detsize}  presents a comparison between the origional CRIRES focal plane array area and the current array of CRIRES$^+$. 
    
    The detector array, composed of three Hawaii 2RG detectors (the CRIRES$^+$ H2RG detectors are shown in Fig.~\ref{fig:det}) span 6144 x 2048 pixels (111 mm x 37 mm) at a pixel size of 18 $\mu m $. For comparison, the old Aladdin mosaic spanned only 4096 x 512 pixels (111 mm x 14 mm) with a pixel size of 27 $\mu m $.
    Also, the gaps between the detectors in the mosaic are smaller. The detectors are operated at 35 Kelvin with cryogenic preamplifiers located next to the focal plane. All detector systems, including the slit viewer camera, were upgraded to the current ESO standard New General Detector Controller NGC \citep{2009Msngr.136...20B}. This detector upgrade not only significantly increased the coverage of the focal plane, but the increased spatial homogeneity of the pixel response, lower readout noise and dark current as well as higher Quantum efficiency result in improved data quality. 

\subsection{Spectropolarimetry with CRIRES$^+$}

    The CRIRES$^+$ instrument has a new spectropolarimetry unit (SPU) which allows one to obtain spectra in circular (Stokes {\textit V}) and linear (Stokes \textit{QU}) polarization in YJHK bands (between $\sim$0.95 and $\sim$2.5 $\mu$m.) The SPU for CRIRES$^+$ uses polarizing gratings (PGs) to split the incoming converging beam into left- and right-circularly-polarized beams which continue along parallel optical axes. This design is quite different from other optical and near-infrared spectropolarimeters such as ESPaDOnS, HARPSpol or SPIRou \citep{2006ASPC..358..362D,2011Msngr.143....7P,2012SPIE.8446E..2EP}. It arises from constraints in the CRIRES$^+$ optical design: mostly the very limited volume, and the location of the SPU being before the adaptive optics in the optical path. This last constraint requires the SPU to transmit optical light essentially unaltered to allow the AO to function, while acting as a beam-splitter in the near-infrared. The choice of PGs as polarizing elements was motivated by their different behaviour at short and long wavelengths, their small thickness, the possibility of producing large and homogeneous samples, and their modest price. The geometry of the periodic pattern that makes up the PGs is chosen such that infrared light (with wavelength longer than $\sim$0.95 $\mu$m) is deviated, while optical light is not. Thus, a pair of PGs act as a polarizing beam splitter for circular polarization without disturbing the operation of the AO system in the optical \citep{2014SPIE.9147E..8PL}. 
    
\begin{figure}
   \centering
   \includegraphics[width=1.\columnwidth]{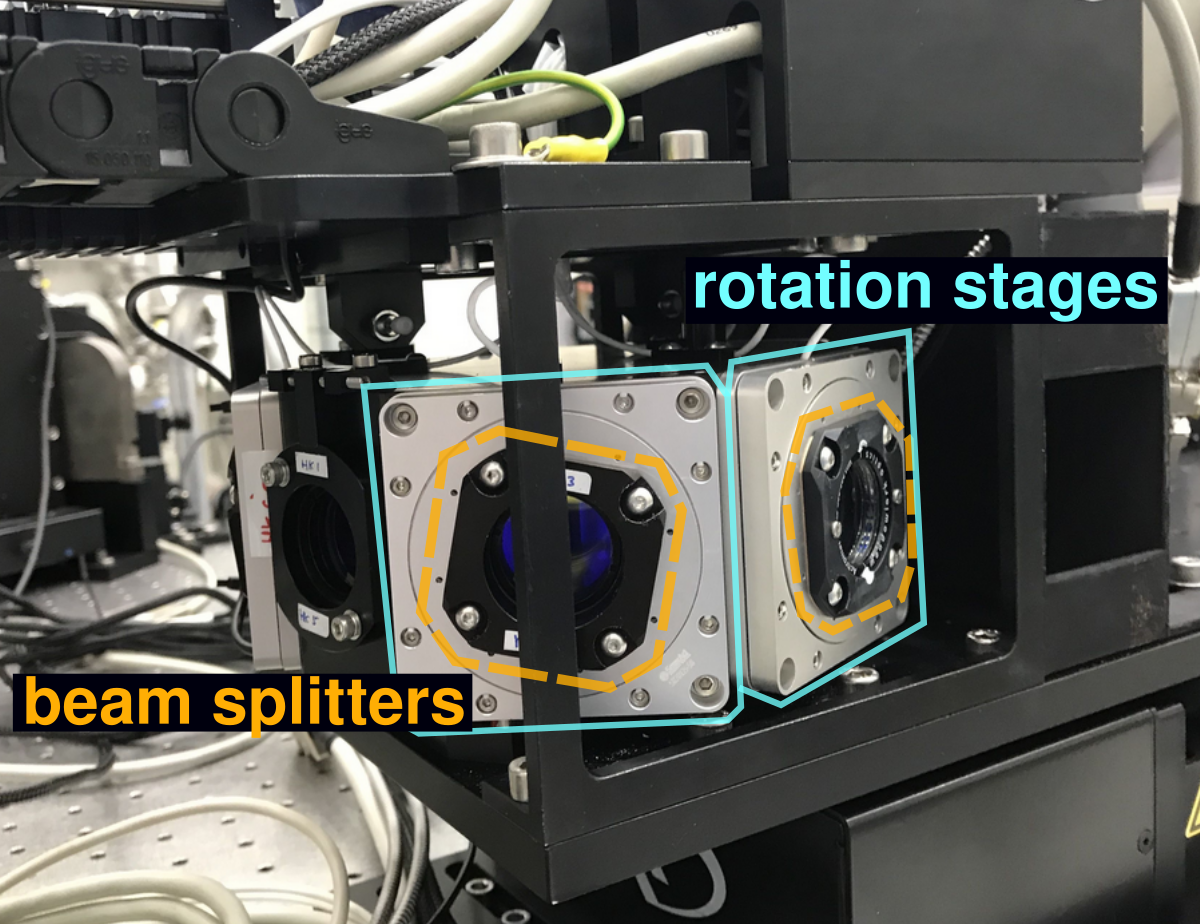}
      \caption{Polarimeter unit mounted on the calibration slide on the warm part of the CRIRES$^+$ instrument. Polarization optics are mounted on the sides of a rotating turret.  }
         \label{fig:spu_figure}
   \end{figure}

    The SPU is very compact and is installed on the CRIRES$^+$ calibration slide as shown in Fig.~\ref{fig:spu_figure}. It is inserted in the optical path when needed. It comprises beam splitters for circular polarization -- consisting of a pair of rotating PGs -- and beam splitters for linear polarization -- consisting of a rotating quarter-wave plate (converting the linear polarization into circular polarization) ahead of a fixed pair of PGs. It was challenging to design beam splitters with sufficient transmission over the entire wavelength range in YJHK bands. We therefore optimized beam splitters for YJ bands and HK bands resulting in four beam splitters: YJ circular, YJ linear, HK circular, and HK linear. The four beam splitters are mounted into a structure atop a rotating stage which selects which beam splitter to place into the optical path. 

    A typical polarimetric exposure sequence consists of four consecutive sub-exposures. In the case of circular polarization (Stokes $V$), the beam splitter rotates between the sub-exposures. Starting from an initial angle (the angle which aligns the two polarized beams on the slit) for the sub-exposure 1, the beam splitter is rotated by 180 degrees for sub-exposures 2 and 3, and rotated back to the original angle for the sub-exposure 4. In the case of linear polarization (Stokes $QU$), the rotating optical element is the quarter-wave plate (QWP) while the beam splitter is fixed. For Stokes $Q$, the QWP is set up for the sub-exposure 1 to an initial angle and then rotated by increment of 90 degrees for sub-exposures 2, 3, and 4 (position angles are 0, 90, 180, and 270 degrees). The sequence for Stokes $U$ is the same than for Stokes $Q$, but with an additional angle of 45 degrees added to each sub-exposures (position angles are 45, 135, 225, and 315 degrees).
    
    The Stokes $IQUV$ spectra are obtained from the sub-exposures using the typical demodulation equations presented in \citet{donati1997MNRAS.291..658D}. This method allows to account for the different optical paths of the two beams through the instrument and remove first-order polarization systematics.

\subsection{The CRIRES$^+$ metrology concept}
    In order to effectively calibrate CRIRES$^+$, the echellogramme alignment on the Focal Plane Array (FPA) needs to be highly repeatable. For example, exoplanetary atmosphere studies require sub-pixel alignment of the echellogrammes. The original CRIRES was limited by a spectral format reproducibility of a few pixels due to imperfect positioning of the echelle grating mechanisms. However, a system of metrology was developed (though never implemented in operations) that facilitated the fine-tuning of the positioning of these mechanisms such that a few tenths of a pixel reproducibility was achieved. Since CRIRES$^+$ retains the original echelle grating mechanism and also the cross dispersion grating wheel did not perform as expected using the locking mechanism, an adapted version of the metrology system was required. 
    
    The new strategy was to centre a defined set of emission lines of the Krypton and Neon pen-ray lamps on the science detector by fine tuning the positions of the cross-disperser grating and Echelle grating and refining further via the use of a piezo driven tip-tilt mirror that has actuators aligned with the main- and cross-dispersion axes. This is an iterative process which may take a few minutes, the exact duration depends upon the unpredictable behaviour of the cross-disperser grating and Echelle grating functions. The metrology then ensures that these emission lines are indeed located at their fiducial positions on the science FPA before any science exposure (or any calibration exposure when used during daytime) follows. Unexpected changes between MAIT testing and installation at the telescope meant that some fundamental modifications needed to be made remotely during the period spanning pre-commissioning to early operations.
    
    Not only were the modifications successful, some improvements were made providing post-metrology stability monitoring, better logging of metrology status in subsequent image headers and the use of better S/N H- and K-band calibration features in L- and M-band metrology \citep{Bristow2022SPIE12184E..5XB}. The sometimes erratic behaviour of the mechanisms means that there are rare occurrences in which the metrology procedure does not converge. Several options are offered to users for how to react in these cases. Independent measurements of echellogramme alignment using Uranium-Neon (UNe) lamp spectra (for main dispersion) and flat fields (for cross-dispersion) confirm that the repeatability goals (<0.2 pixel main dispersion, <0.5 pixel cross dispersion) are achieved. The pen-rays used for the metrology have spectral features with a high dynamic range, some are saturated in the exposure times used (a few to a few tens of seconds) while many more have very low signal to noise, neither case yields a precise centroid. Consequently a few lines in each setting with high signal to noise, but with all pixels comfortably below the saturation level, are hand-picked for use in the metrology.

\subsection{New data reduction software}
    The CRIRES$^+$ project provides the community with a new data reduction software (DRS) package which supports all of the offered observing modes, and makes use of the new calibrations. The improved repeatability of the instrument settings and the newly defined set of standard echelle angle settings allow re-use of calibration data between nights, thereby minimizing overheads. In practice, however, most calibrations are taken during the day after the observations. 
    
    The pipeline corrects detector effects (dark, flat-field, non-linearity) and determines the location of the spectral orders in the detector frame, as well as the tilt of the slit within each order. This information is used to extract the spectra optimally to render the best signal-to-noise ratio \citep{2021A&A...646A..32P}. For the new polarization mode, sets of frames with different rotation angles of the beam splitter are combined and demodulated to determine the Stokes parameters on the pixel-level, before extraction. Instrument and sky background is subtracted both via nodding or sky-frames, and by using the inter-order gaps on the detectors. The wavelength calibration in Y, J, H, and K band uses carefully selected lines from the UNe lamp for zero-pointing, and a 2D fit to the Fabry-Perot etalon fringes to achieve good accuracy. 
    
    Non-simultaneous absolute wavelength calibration accuracy is estimated to be below 300 m~s$^{-1}$ in Y, J, and H-bands and 500 m~s$^{-1}$ in the K-band due to the scarcity of UNe calibration features. In the L and M bands it is dependent upon applying interactively derived static solutions and therefore depends upon the reproducibility that in turn depends upon the metrology. In practise this can be assumed to be well below 1000 m~s$^{-1}$. Use of the Fabry-Perot interferometer permits non-simultaneous relative wavelength calibration accuracy in YJHK bands of <30 m~s$^{-1}$. In L and M bands it is likely to be worse and rather dependent upon the lines available to the static solutions.
    
    The use of molecfit, a tool to correct for telluric absorption lines based on synthetic modeling of the Earth’s atmospheric transmission \citep{2015A&A...576A..77S}, is currently under development for CRIRES$^+$. This will facilitate improved (both better accuracy and better characterisation of the accuracy) absolute and relative wavelength calibration in the L and M-bands.

\subsection{Warm part and MACAO refurbishment and recovery }
    The foreseen lifetime for the upgraded CRIRES$^+$ is at least ten years. CRIRES was operated in conjunction with a 60-element curvature adaptive optics system, Multi-Application Curvature Adaptive Optics (MACAO), described by \citep{2004SPIE.5490..216P}, and required interventions to prevent its obsolescence. This was already planned for the MACAO Very Large Telescope Interferometer (VLTI) systems installed in the Coudé laboratory of the VLT Unit Telescopes (UTs). Accordingly, the CRIRES MACAO system was refurbished in a similar manner to the VLTI systems by replacing and upgrading obsolete electronic boards. In addition, it was planned to exchange the membrane mirror, re-coat additional mirrors, realign the optics and re-commission the full AO system. 
    
\begin{figure*}
   \centering
   \includegraphics[width=15cm]{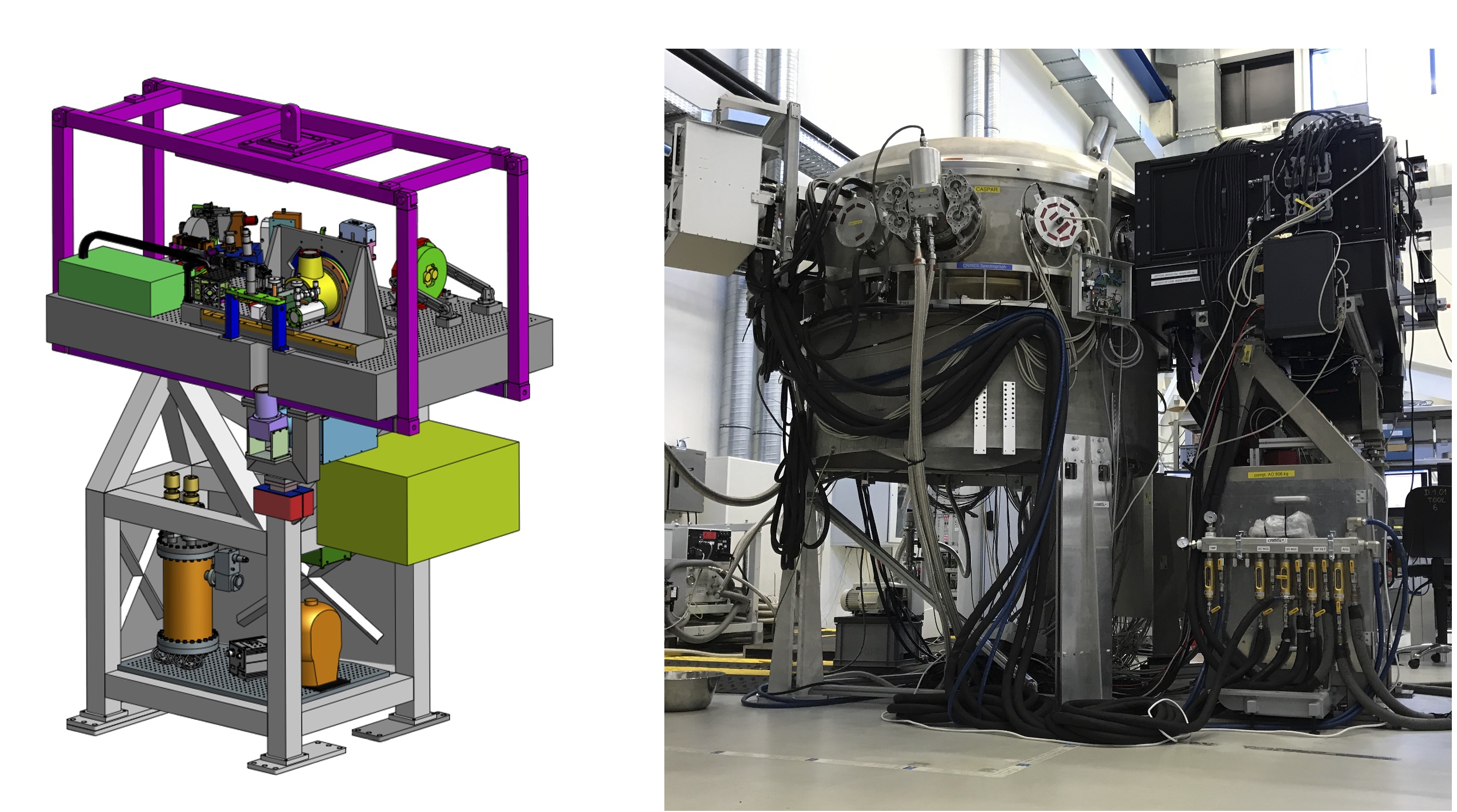}
      \caption{Left: New CRIRES$^+$ warm part assembly with etalon system, calibration slide, AO system and de-rotator mechanism.  Right: Refurbished and enhanced warm part including the AO system mounted to the CRIRES$^+$ cold part during system testing at the ESO integration hall in Garching, Germany. 
              }
         \label{fig:warmpart}
   \end{figure*}

    In July 2014, the MACAO-CRIRES warm optics bench was damaged during its transport. The recovery from this damage then became part of the CRIRES$^+$ project. The warm optics bench was replaced and re-designed to allow better handling and secure installation at the telescope with a crane-lifting device (violet parts of Fig.~\ref{fig:warmpart}). In addition, a loss in throughput of the original CRIRES instrument had been observed over the years, originating from a degradation of the optical quality of the warm mirrors. Thus, the CRIRES$^+$ project aimed at making sure that the optical quality of the warm mirrors did not degrade over the coming years of operation. To measure the mirror quality, the Physikalisch Technische Bundesanstalt Berlin was contracted to characterize all mirrors on the warm part including the Deformable Mirrror (DM). PTB measured the mirrors in the wavelength range of 950 to 5300~nm and at optical wavelength from 400 to 1000~nm. All common path mirror were then replaced with new ones. 

\section{CRIRES$^+$ Instrument performance}
This section summarizes the main characteristics of the CRIRES$^+$ instrument and its performance as measured during commissioning and early science operations. More details and updates are available through the instrument web pages\footnote{https://www.eso.org/sci/facilities/paranal/instruments/crires.html} at ESO.
\subsection{Adaptive optics performance }

\begin{figure}
   \centering
   \includegraphics[width=1\columnwidth]{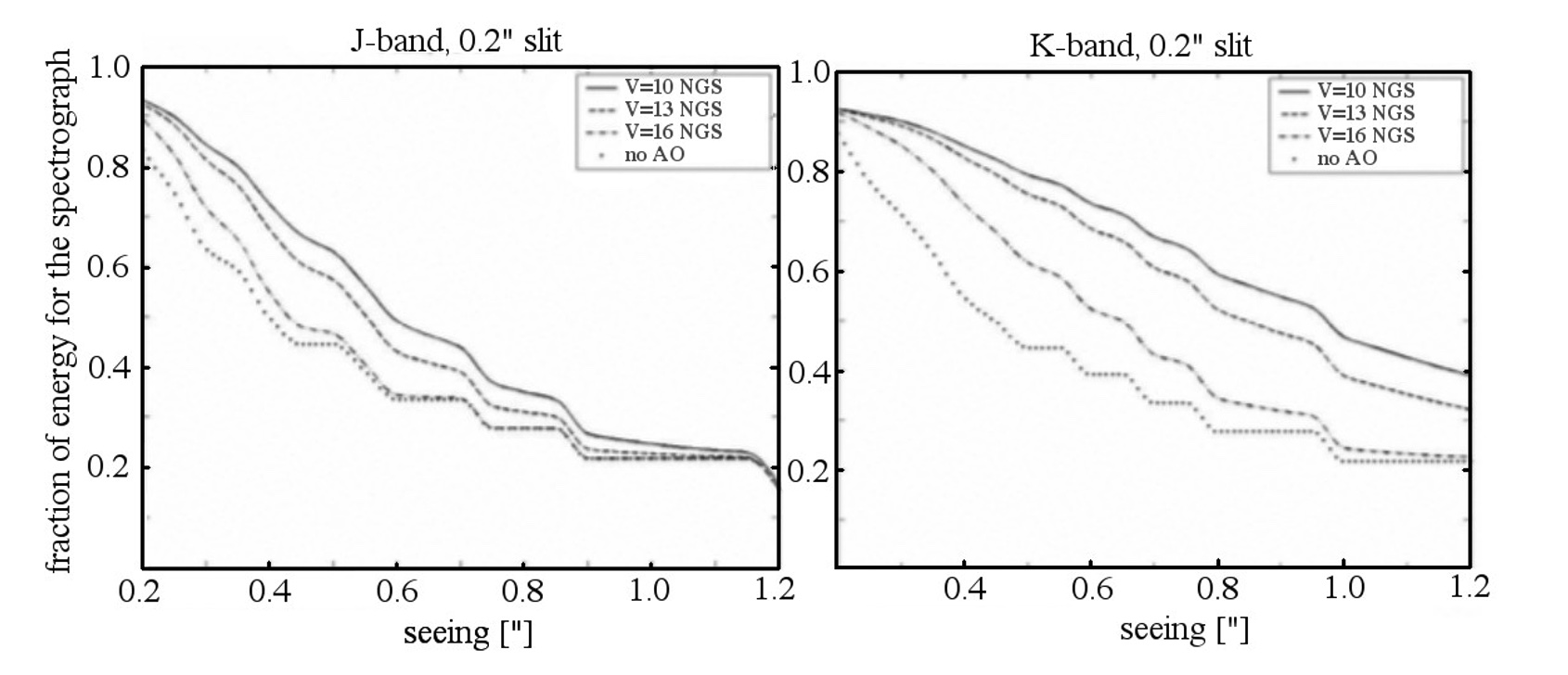}
      \caption{The fraction of energy available for the spectrograph in a 0.2" slit as a function of the optical seeing is shown for the J (left) and K (right) band for an AO natural guide star (NGS) of V=10, 12, 16 mag and without AO correction.
              }
         \label{fig:AOenergy}
   \end{figure}
   \begin{figure}
   \centering
   \includegraphics[width=1\columnwidth]{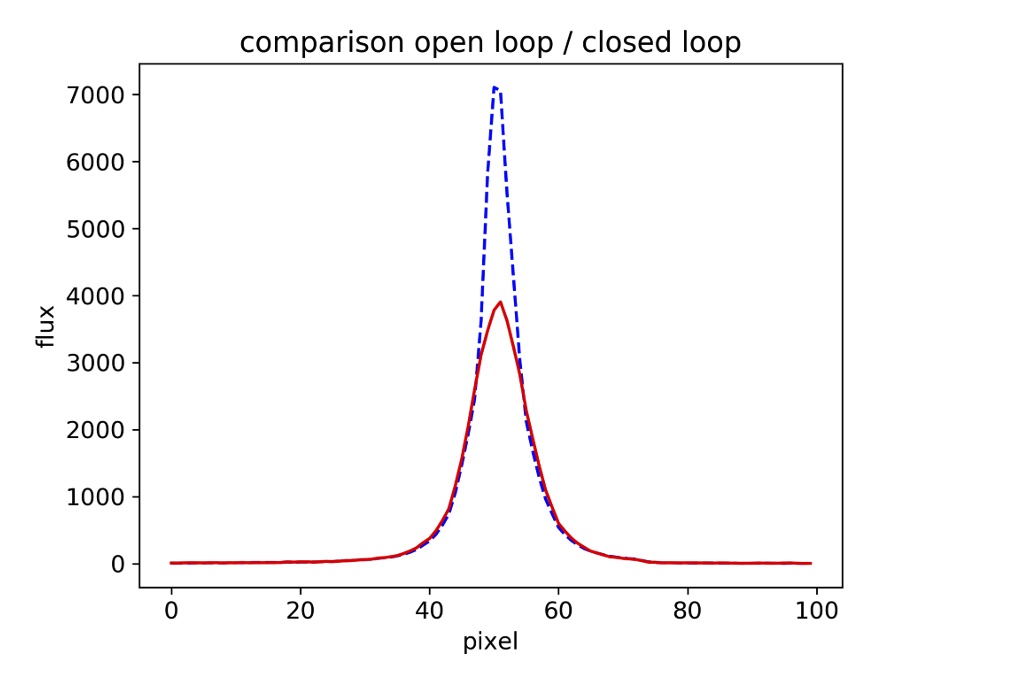}
      \caption{Factor 1.8 adaptive optics improved throughput. H-band flux of the star Pi.02 Ori measured along the 0.4” slit in open loop (solid red line) and closed loop (dashed blue line).
              }
         \label{fig:AOloop}
   \end{figure}
    The performance achieved by the refurbished MACAO system of CRIRES$^+$ has been evaluated by laboratory simulations comparing two cases, in closed loop with guide star of various magnitudes and in open loop (without AO correction). The optimization was done over the encircled energy on a 0.2” slit, representative of the available energy of the spectrograph. Lab results were confirmed by on-sky measurements and demonstrate some gain in the J band (more than 40\% for an optical seeing of 0.6”) and a large increase (factor of $\sim$2) of the fraction of energy available for the spectrometer in K and M band, respectively (see Fig.~\ref{fig:AOenergy}). For seeing $>$1.4”, the AO correction becomes very poor and unstable and does not result in any improvement with respect to the non-AO mode. Fig.~\ref{fig:AOloop} illustrates the increased throughput when the AO system is employed. The graph shows the spatial profiles of the spectrophotometric star Pi.02 Ori (R=4.29) at a wavelength of 1559.245~nm taken in atmospheric condition corresponding to Turbulence category = 50\%, in open and closed loop. In both cases, the exposure times were the same; however, in closed loop a flux level about 1.8 times higher was attained than in the open loop observations.

\subsection{Detector characteristics and detector readout-mode}
   \begin{figure}
   \centering
   \includegraphics[width=1\columnwidth]{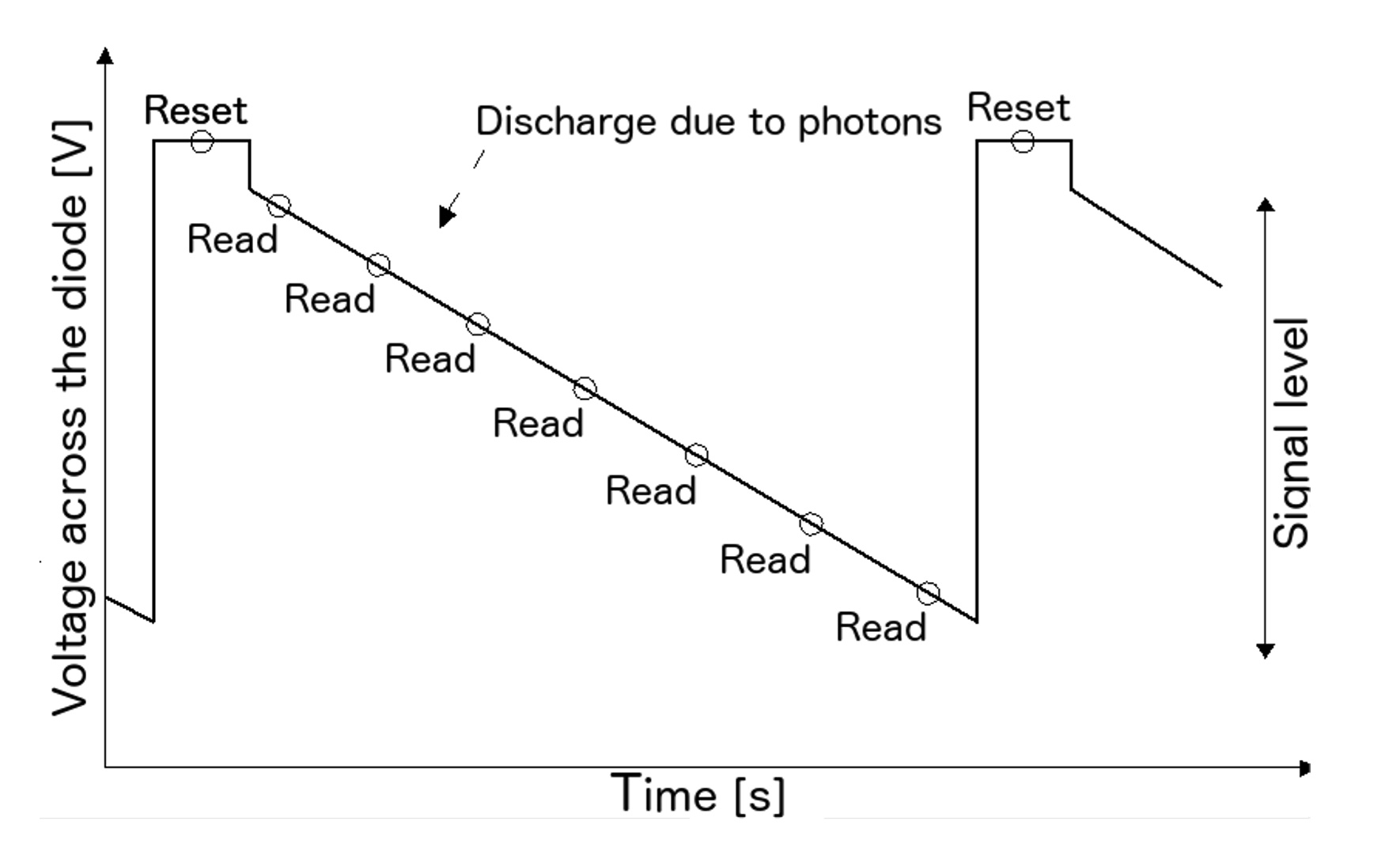}
      \caption{Sample Up The Ramp readout mode. Before each integration, the pixels are reset to the initial capacity. During the integration, the detector is non-destructively read (from two readings in case of the minimum DIT up to a maximum of 36 readings for long DITs). These detector readings are equidistantly spaced in time. The flux rate per pixel corresponds to the slope of the flux values of the subsequent readings.
              }
         \label{fig:readoutmode}
   \end{figure}
   
   \begin{figure}
   \centering
   \includegraphics[width=1\columnwidth]{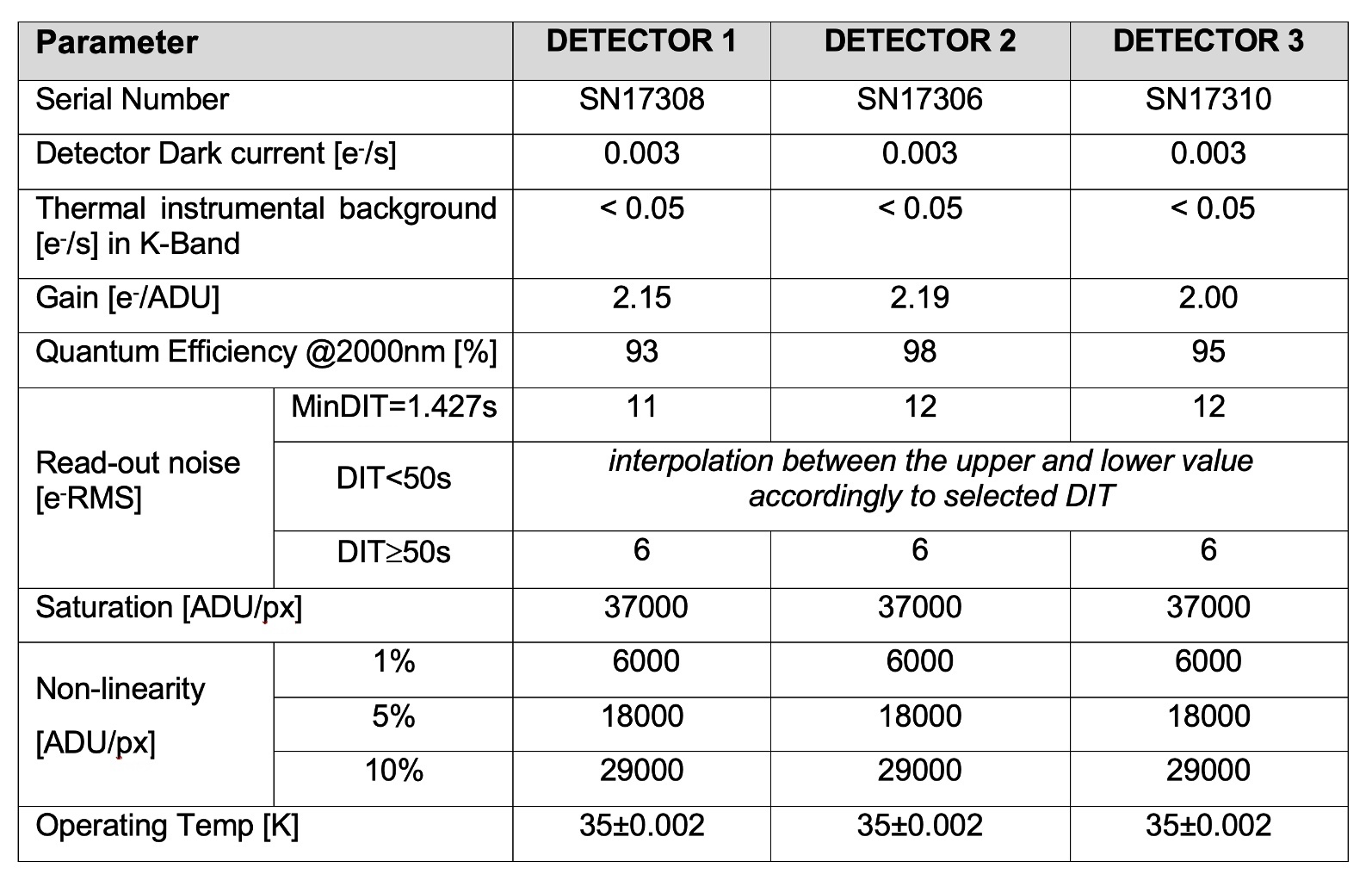}
      \caption{Main performance parameter for the scientific detector system. 
              }
         \label{fig:dettable}
   \end{figure}
   
\begin{figure}
   \centering
   \includegraphics[width=1\columnwidth]{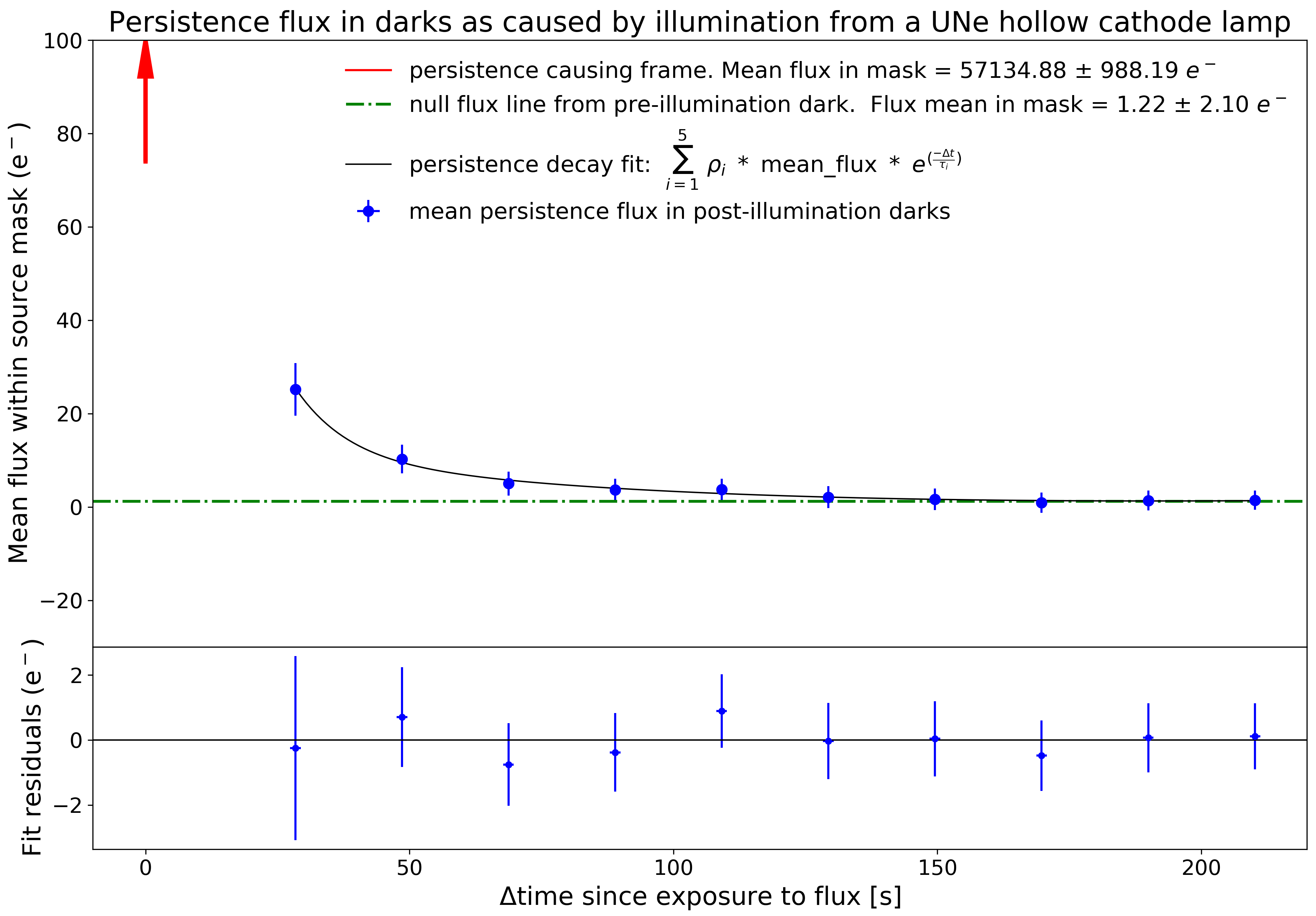}
      \caption {Persistence decay as a function of time since exposure to emission line flux for detector one. Following an exposure with the Uranium Neon hollow cathode calibration lamp, in which the brightest spectral line reached almost 60k electrons (indicated by the red arrow), we mapped the persistence in a series of ten, post-illumination darks (blue points). Plotting the brightest feature, it is evident that the persistence signal decays to the pre-illumination dark level in less than 200 s.  A similar decay rate is observed for the other two detectors.
              }
         \label{fig:plot_persistence}
   \end{figure}

    The main performance parameter for the scientific detector system are shown in Fig.~\ref{fig:dettable}. The only detector readout mode provided for CRIRES$^+$ is Sample Up The Ramp (Fig.~\ref{fig:readoutmode}). The exposure time is set by the detector integration time (DIT) and the number of such integrations (NDIT) to be averaged into one single exposure, whose total integration time is therefore NDIT$\times$DIT. The minimum DIT for full frame readout is 1.427 s (Minimum two samples). The optimum combination of DIT and NDIT is determined by the CRIRES$^+$ Exposure Time Calculator (ETC)
    \footnote{https://etc.eso.org/observing/etc/}. 
    The CRIRES$^+$ ETC web application has been available since March 2021. Utilities to call the ETC backend calculation engine non-interactively through an API (application programming interface), from a command line or script, are available in the 'Tools' tab  in the ETC 2.0 web application. Bright objects or observations in the L or M bands (high thermal background) require short DITs to stay below saturation due to the thermal background, as heavily saturated pixels lead to detector persistence that require a couple of minutes to decay to the extent that it does not affect subsequent exposures anymore. H2RG detectors are known to have significant persistence and before integration in the instrument, the CRIRES$^+$ detectors  were characterized in the laboratory. Measurements during commissioning indicated that while the specification that the signal persistence five minutes after an exposure to 90\% full well shall not exceed 1\% (goal: 0.5\%) is clearly satisfied, there was also evidence that persistence was detectable in darks obtained directly after wavelength calibrations. As a mitigation 'cleaning darks' were introduced following wavelength calibrations. These were effective in removing the persistence signal in the real darks while also providing a means to monitor persistence decay as shown in Fig.~\ref{fig:plot_persistence}.
    
\subsection{Instrumental background and detector dark current}
\begin{figure}
   \centering
   \includegraphics[width=1\columnwidth]{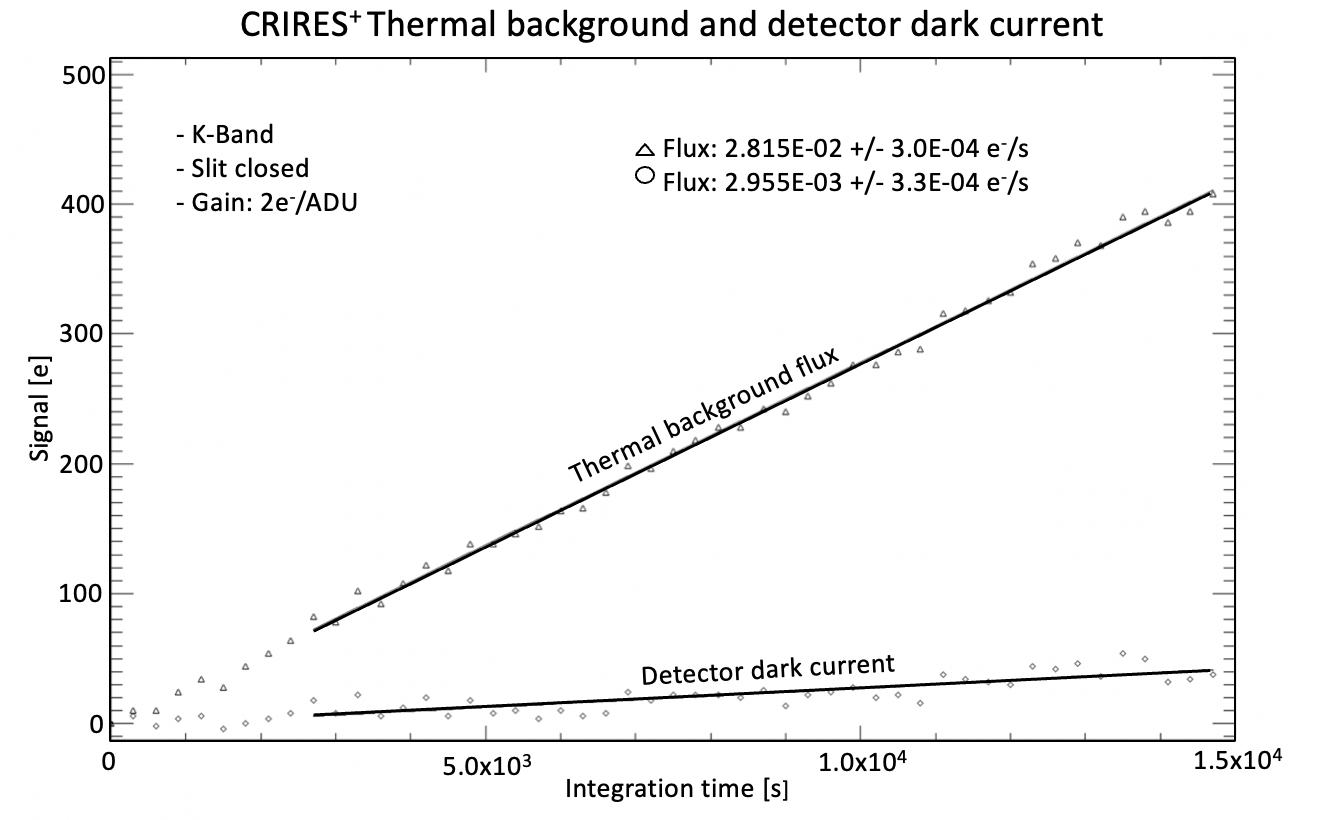}
      \caption{Dark current of the detectors compared to the thermal background of the instrument in K-Band on scientific detector number 3.
              }
         \label{fig:background}
   \end{figure}
    The dark current and instrumental background was estimated from the slope of the signal as a function of the integration time (s) for the linear region. Fig.~\ref{fig:background} shows a dark current of 0.003 electrons/pixel of detector number~3 measured in K-band on a covered area of the detector to exclude the thermal background of the instrument. The thermal background with the entrance slit closed is of the order of 0.03 electrons/pixel, making the instrument about 33 times darker than the original CRIRES before the upgrade.  However, with the slit open we measure a higher background in L and M band compared to the original CRIRES which is under investigation. A possible reason could be the overall higher throughput of the instrument. 
    
\subsection{Pipeline correction for detectors non-linearity}
\begin{figure}
   \centering
   \includegraphics[width=1\columnwidth]{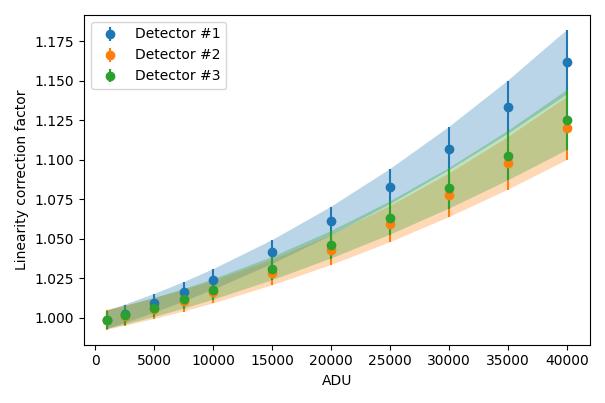}
      \caption{Correction for non-linearity effects implemented by the pipeline. The correction factor from the polynomial fits (pixel-by-pixel), evaluated at certain ADU levels and plotted as medians and 1-sigma shaded regions over all pixels is shown.
      }
    \label{fig:lin1}
   \end{figure}
   
\begin{figure}
   \centering
   \includegraphics[width=1\columnwidth]{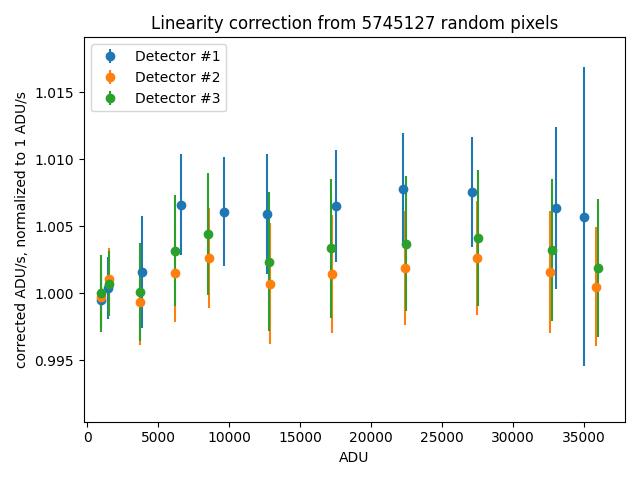}
      \caption{Normalized ADU/s as measured from the frames corrected for non-linearity. For the three detectors, the median over the bins in ADU-level is given. A value of 1.005 on the y-axis corresponds to 0.5\% inaccuracy.
              }
         \label{fig:lin2}
   \end{figure}

    All common IR detectors suffer from non-linearity effects. In the case of CRIRES$^+$, deviation from linearity is of the order of 5\% of the detected flux at about 18k ADUs and increases with flux. However, the CRIRES$^+$ pipeline can correct for non-linearity effects at low-, medium- and high- count levels as illustrated in Fig.~\ref{fig:lin1} and Fig.~\ref{fig:lin2}.

\subsection{Slit viewing camera}
\begin{figure}
   \centering
   \includegraphics[width=1\columnwidth]{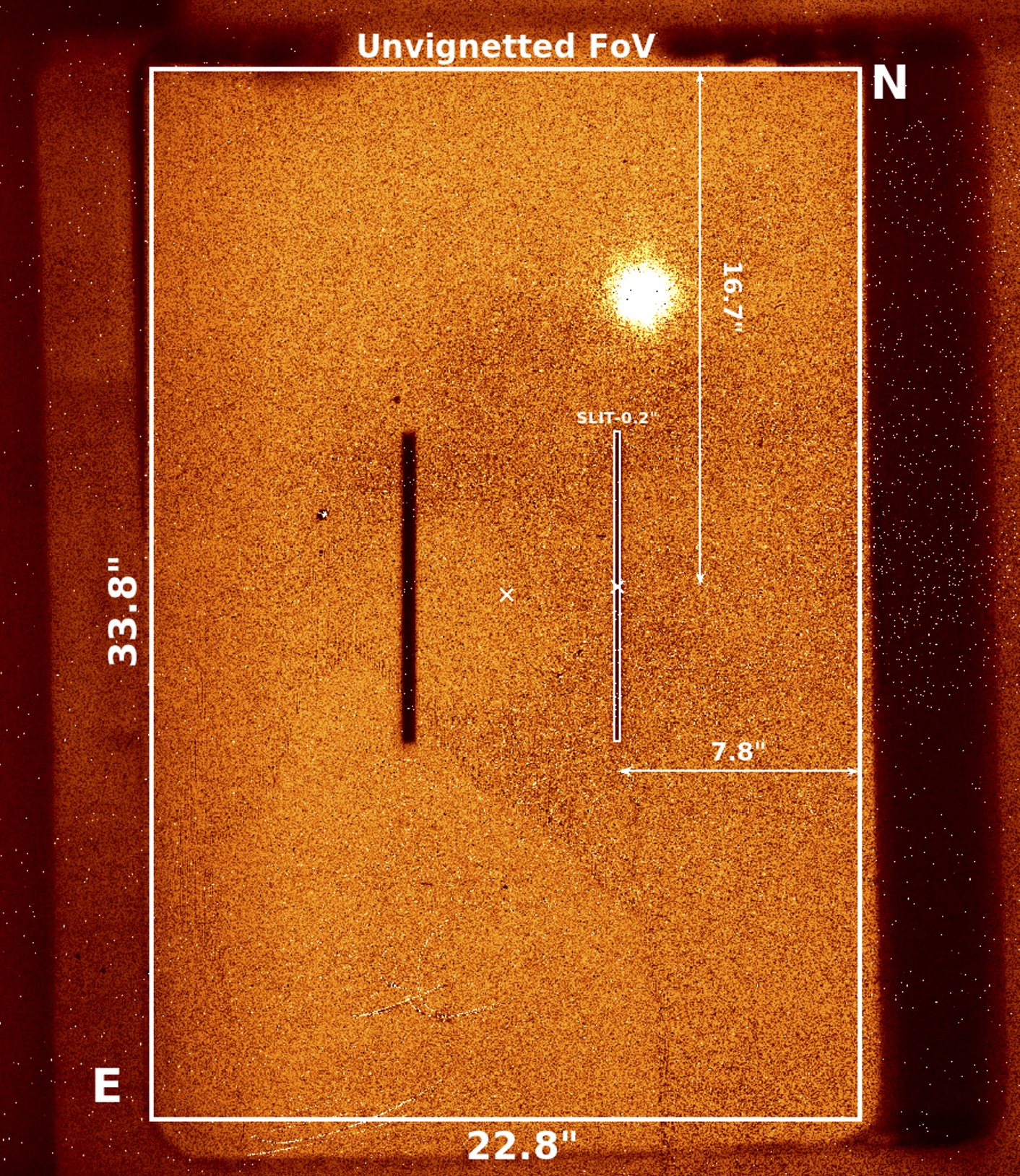}
      \caption{Geometry of the SV detector for a position angle of 0 deg. on sky. North is up and east is on the left. The unvignetted FoV usable for target acquisition and guiding is around 22.8” x 30.8”. The centre of the usable FoV and of the slit is marked as white cross. The slit centre (0.2” in this example) does not match the unvignetted FoV centre, but it’s displaced by 3.6” W and 0.2” N. When using the 0.2” slit, the footprint of the 0.4” slit vignettes an area of 0.4”x10” located 3.6” eastwards with respect to the FoV centre (see vertical dark stripe). When the 0.4” slit is used, there is no vignetting from the 0.2” slit as its footprint falls at the border of the FoV.}
         \label{fig:slit}
   \end{figure}
Target centring and natural guide star acquisition are performed in the infrared via the CRIRES$^+$ Slit Viewer camera (SV). Because CRIRES$^+$ observations require the use of a guide star (SVGS) to ensure that the target is properly kept centred on the slit during the science exposures, the SVGS is acquired through the SV camera. With a pixel scale of 37.3~mas, the SV covers a maximum unvignetted sky projected field of view (FoV) of 22.8’’$\times$33.8’’, with 33.8’’ along the slit and 22.8’’ perpendicular to the slit, thus making the available FoV slightly smaller than that of the old CRIRES instrument. The position of the slit within this window is off-centred by 3.6’’ towards the west, and 0.2’’ northwards, as illustrated in Fig.~\ref{fig:slit} , to increase the allowed maximum separation between the target and the SVGS.

When the target is also used as SVGS, guiding is performed using the light reflected off the slit viewer window around the slit.  CRIRES$^+$ also utilizes a de-rotator to control the alignment of the slit relative to the sky (namely to compensate for field rotation in the Nasmyth focus) or to align the slit with the parallactic angle so as to reduce slit losses. Finally, it is worth noticing that when the 0.2” slit is used, the footprint of the 0.4” slit vignettes an area of 0.4’’$\times$10’’ about 8’’ eastwards with respect to the slit centre. The SV is equipped with six NIR filters: J, H, K, two neutral density H filters and one neutral density K filter. The SV is sufficiently sensitive that any emitting point source for which one aims to obtain a spectrum should be seen on the SV image. During the on-sky commissioning run, under excellent conditions (0.5” seeing) and a slit of 0.2”, reasonable guiding was possible with stars of H=14.5 and 13.5 in Non-AO and AO mode, respectively. 

\subsection{Performance of the spectropolarimetry unit}
\begin{figure}
   \centering
   \includegraphics[width=1.\columnwidth]{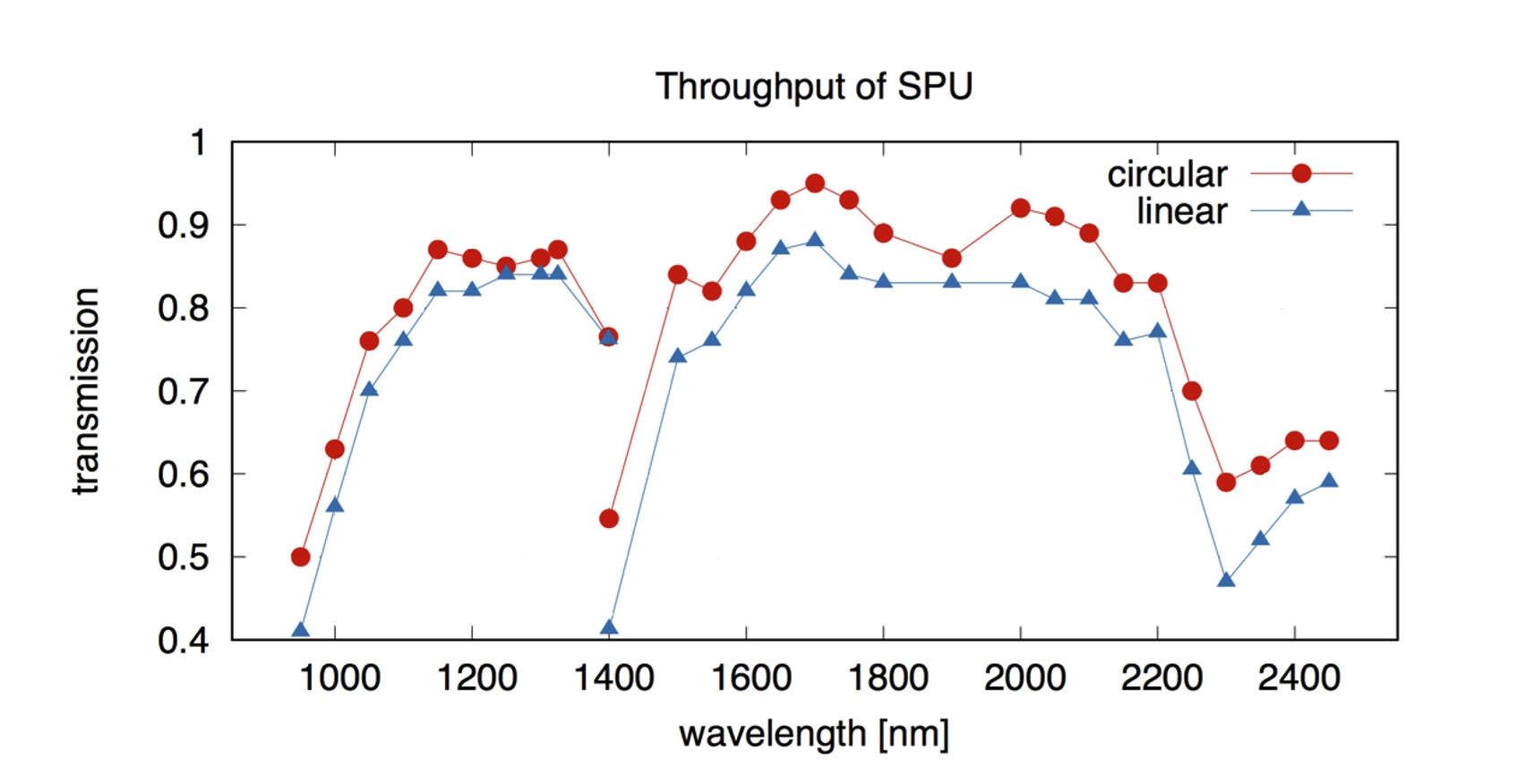}
      \caption{Measurements of the throughput of the spectropolarimetry Unit (SPU).
              }
         \label{fig:spu}
   \end{figure}
\begin{figure}
   \centering
\includegraphics[width=1.\columnwidth]{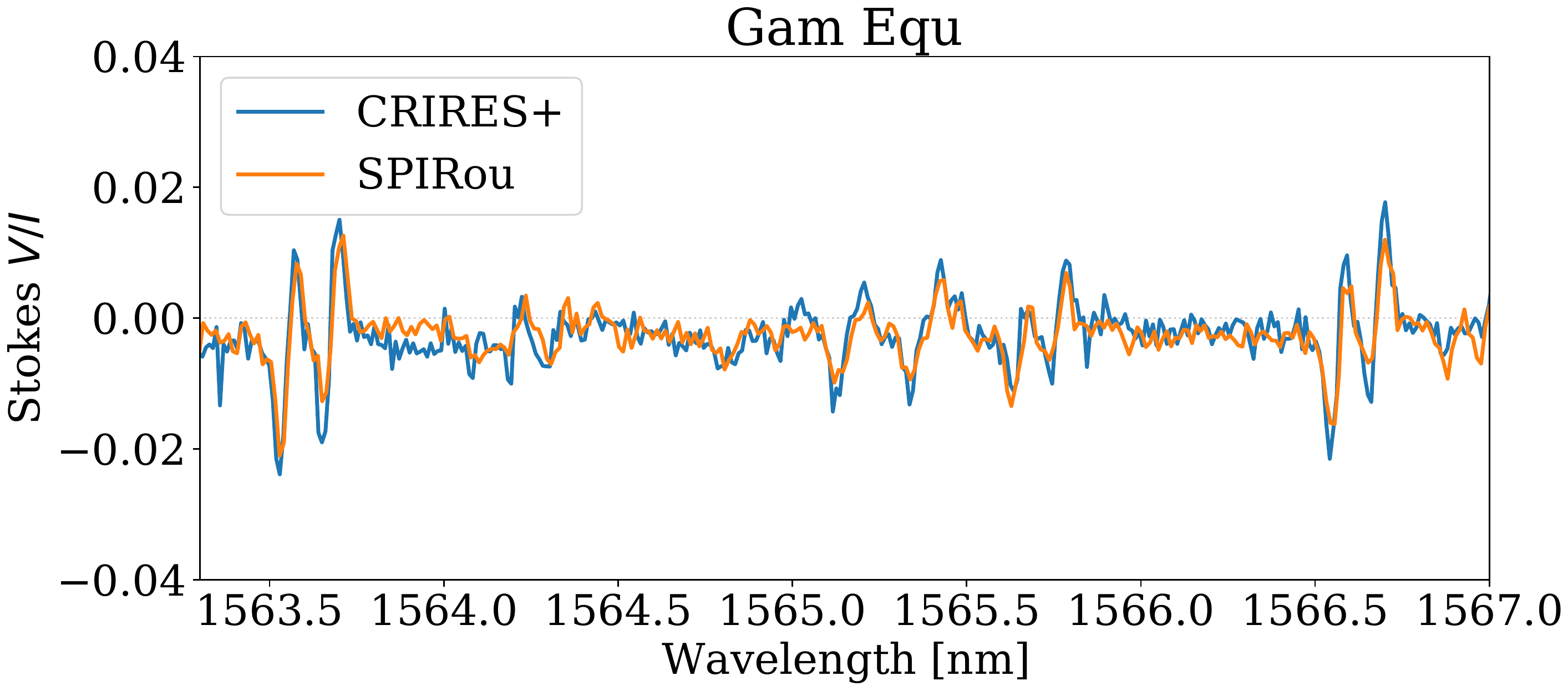}
      \caption{Stokes~$V$ (circularly polarized) spectra of the star Gam Equ observed with CRIRES$^+$ (blue line) and SPIRou (yellow line) in the H band.
              }
         \label{fig:crires-spirou-pol}
   \end{figure}

    Fig.~\ref{fig:spu} shows the throughput of the SPU, respectively for circular and linear polarization. A comparison with the high-resolution near-infrared spectropolarimeter SPIRou working in the same spectral region was done using the spectropolarimetric standard star Gamma Equ. The polarization signal was compared in Y to K bands for selected lines with strong Zeeman effect. We find good match for Stokes V and Q profiles. Using the intensity profiles, it was possible to confirm that the residual differences are due to higher spectral resolution of CRIRES$^+$. An example of this comparison is shown in Fig.~\ref{fig:crires-spirou-pol}. The adaptive optics system components (deformable mirror, wavefront sensor) are both located after the polarimeter in the optical path of the instrument and do not affect polarimetry. The AO system improves the resulting signal to noise and for a given exposure time, higher polarimetry sensitivity and precision are achieved.

\subsection{Characteristics of the spectrograph }

\begin{figure}
   \centering
   \includegraphics[width=1\columnwidth]{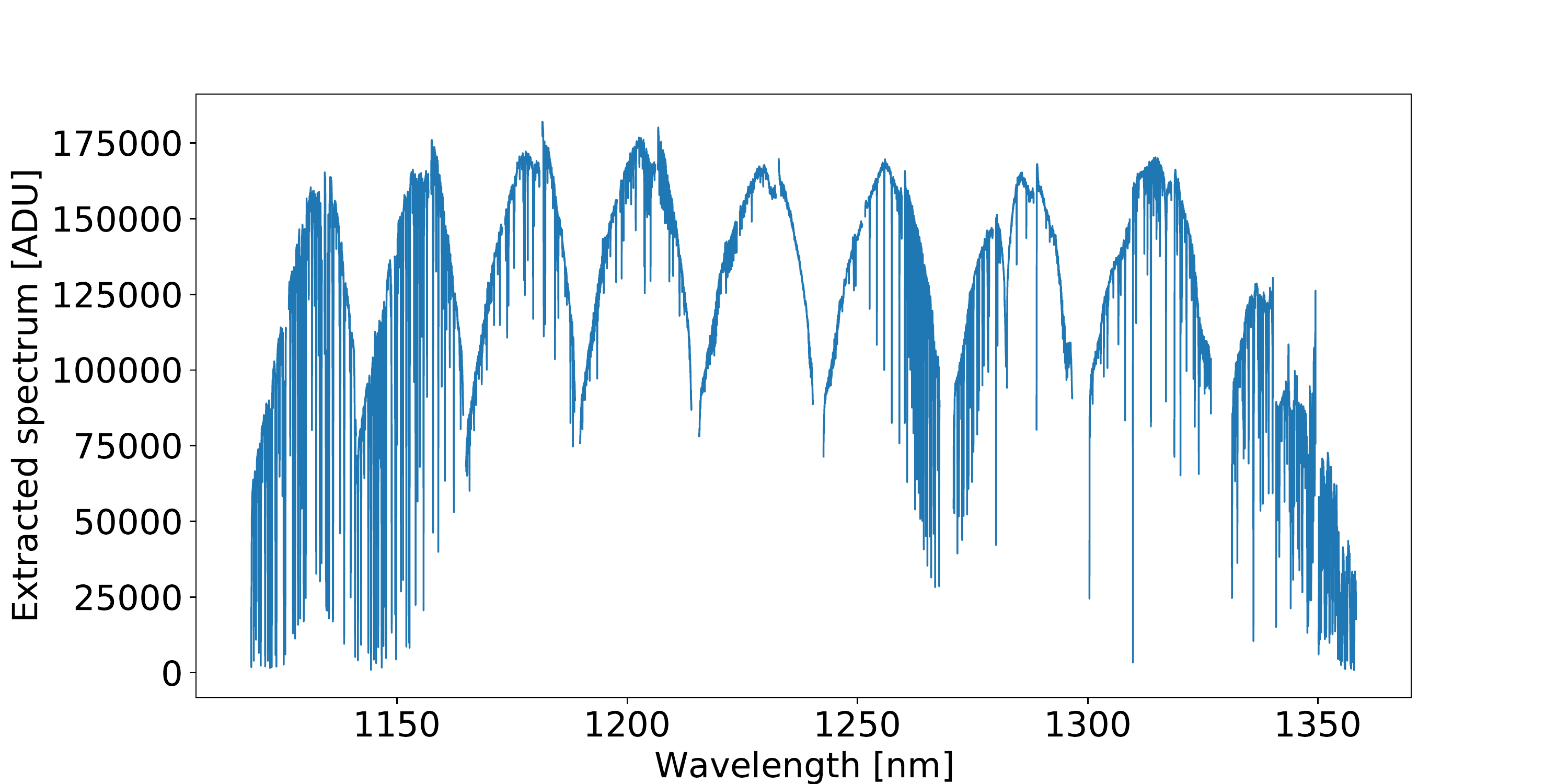}
      \caption{1D extraction of a stellar spectrum  indicating the spectral coverage achieved for a single wavelength setting, here J1228.
              }
         \label{fig:1D}
   \end{figure}

\begin{figure*}
   \centering
   \includegraphics[width=1.7\columnwidth]{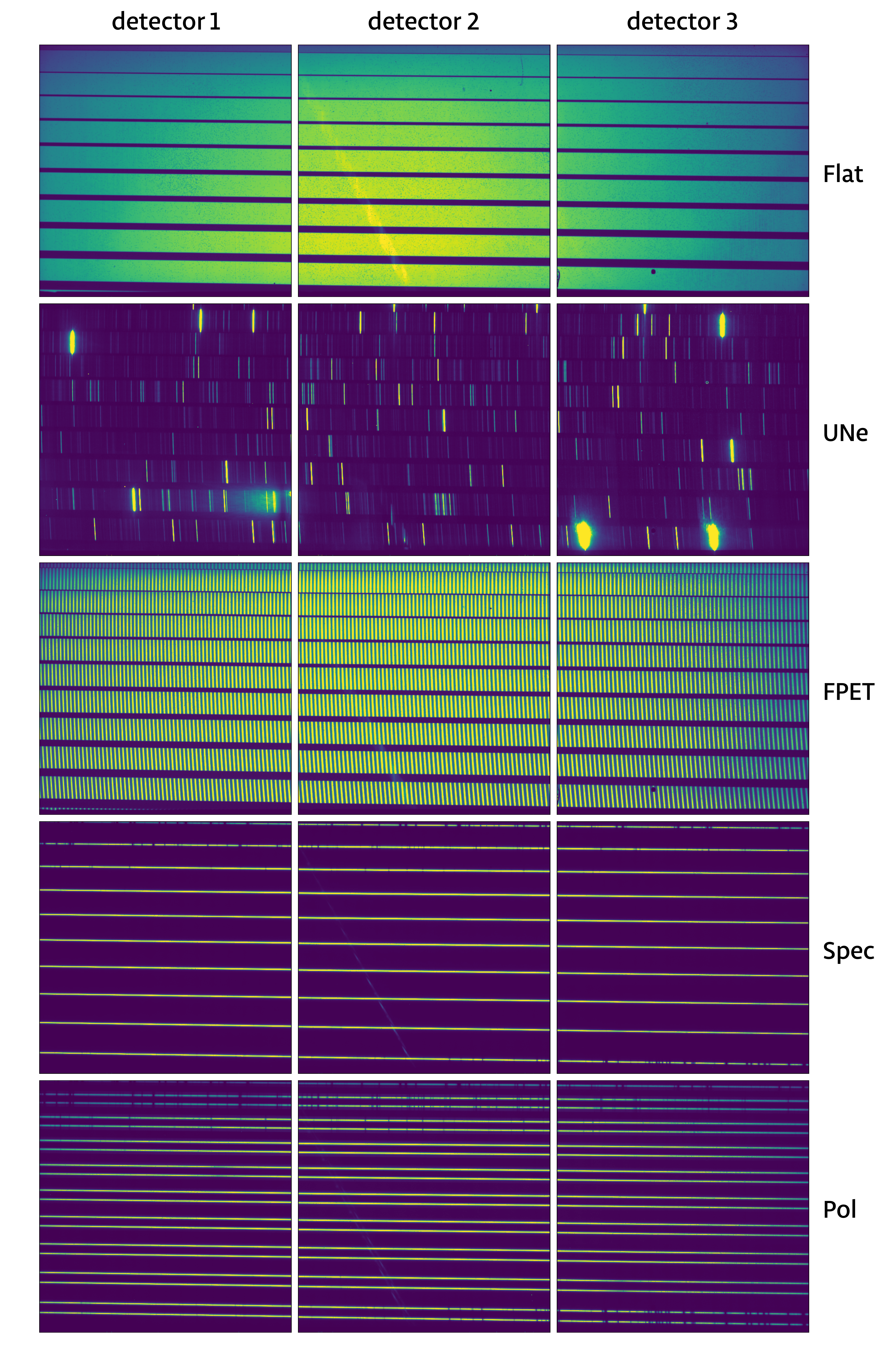}
      \caption{Each row shows a different type of CRIRES$^+$ raw frame acquired with its three science detectors, all taken in the wavelength setting Y1029. Bad pixels have been masked. Starting from the top, the rows show: a flat-field exposure (Flat), a wavelength calibration frame with the Uranium-Neon lamp (UNe), a wavelength calibration frame taken with the Fabry-Perot etalon (FPET), a stellar spectrum (Spec), and a stellar spectrum taken with the spectropolarimeter unit in the optical path -- resulting in two spectra with opposite polarization being recorded (Pol). We note in passing that on detector~2, an internal reflection (Littrow ghost) produces a faint, diagonal strip. 
              }
         \label{fig:raw-frames}
   \end{figure*}

    The introduction of the cross-disperser increases the single-exposure wavelength coverage of CRIRES$^+$ up to a factor of 10 compared to the original CRIRES, but CRIRES$^+$ is still incapable of covering a single photometric band in a single exposure without any gaps. By varying the echelle angle and choice of cross-disperser grating, CRIRES$^+$ is able to fully cover each Y, J, H, K, L and M photometric band. The number of exposures depends on the particular band, fewer exposures are required to cover the shorter wavelength regions. Additional exposures are necessary to cover detector gaps. Similar to the standard settings offered with the original CRIRES, CRIRES$^+$ provides a list of fixed wavelength settings to the users. All settings are handled by the Data Reduction Software (DRS). An extracted stellar spectrum as output from the DRS for an example wavelength setting in the J band is shown in Fig.~\ref{fig:1D}.  Examples of different types of CRIRES$^+$ raw frames are shown in Fig.~\ref{fig:raw-frames} and illustrate the new cross-dispersed spectral format on the detector mosaic. It clearly shows some features arising from the optical design, for example that spectral orders are tilted on the detector mosaic, that the tilt of spectral lines (slit-tilt) is variable as a function of wavelength, and that there is a ghost crossing several spectral orders (visible in for instance the flat-field on detector 2).

    To reduce the total number of fixed settings offered by CRIRES$^+$, the number of settings per photometric band were optimized to provide the best overall throughput per band (in individual and combined images), with the least number of echelle settings needed to provide for gap-free coverage. One of the goals of the CRIRES upgrade was to achieve a minimum of 80\% coverage of the photometric band in the region of operation. In most cases, the expected coverage is much wider. However, as the height of some orders may not be fully covered, the spectral coverage for point sources (at the centre of the slit) differs from the spectral coverage achieved with extended sources that require full slit illumination. Table \ref{tab2} provides a list of the achieved coverage (using multiple exposures) for each photometric band. A total of 29 different wavelength settings is required to cover the full operating range of CRIRES$^+$. Some small wavelength gaps cannot be probed with CRIRES$^+$ due to design decisions to optimize the throughput in the regions of interest. The ranges include the following: 1356-1423 nm, 1854-1908 nm, and 2527-2725 nm (the gaps are larger if full slit illumination is considered). These regions are dominated by telluric lines and are not of general interest for most science cases.
    
     \begin{table*}
      \caption[]{Approximate spectral coverage achieved within different photometric bands.}
      \label{tab2}
\small
\begin{tabular}{p{1.5cm} p{2cm} p{2cm} p{2cm} p{2cm} p{2cm} p{2cm}} 
 \hline
 &\multicolumn{2}{p{5cm}}{\textbf{Spectral coverage of photometric band}} &\multicolumn{2}{p{5cm}}{\textbf{Spectral coverage of point source (middle of slit)}}&\multicolumn{2}{p{5cm}}{\textbf{Spectral coverage for full slit}}\\
  \hline
\textbf{Band} & Starting $\lambda$ (nm)&	Ending $\lambda$ (nm)&	Starting $\lambda$ (nm)&	Ending $\lambda$ (nm)&	Starting $\lambda$ (nm)&	Ending  $\lambda$(nm)\\ 
 \hline
Y&950&1120&948&1120&948&1120\\
 \hline
J&1100&	1400&	1116&	1356&	1116&	1331*\\
\hline
H&1500&	1800&	1423&	1854&	1461&	1796\\
\hline
K&2000&	2400&	1908&	2527&	1946&	2472**\\
\hline
L&3200&	3700&	2810&	4150&	2840&	4100\\
\hline
M&4600&	5000&	3340&	5800***&	3360&	5600***\\
\hline
\multicolumn{7}{c}{*1356 nm 85\% of slit; **2501 nm 75\% of slit; *** The detector cutoff is at 5300 nm}\\
\end{tabular}
   \end{table*}

\subsection{Throughput of the Instrument }
\begin{figure}
   \centering
   \includegraphics[width=1.\columnwidth]{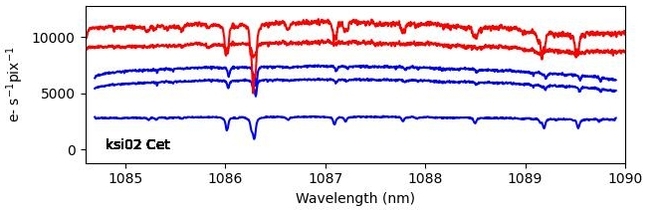}
   \includegraphics[width=0.99\columnwidth]{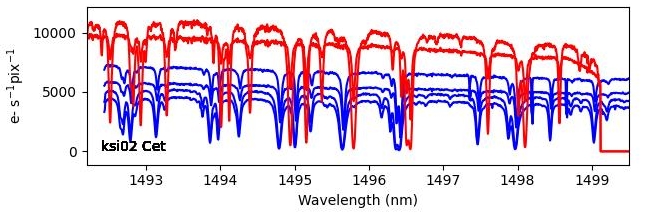}
   \includegraphics[width=1.\columnwidth]{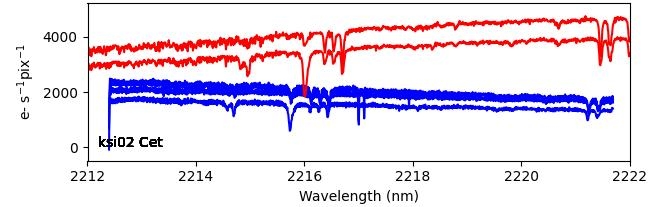}
   \includegraphics[width=0.98\columnwidth]{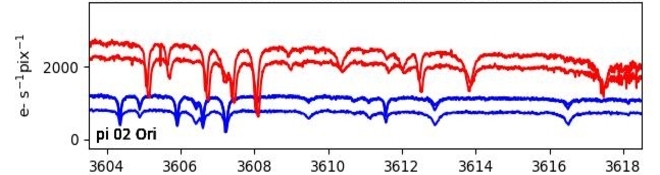}
   \includegraphics[width=1.\columnwidth]{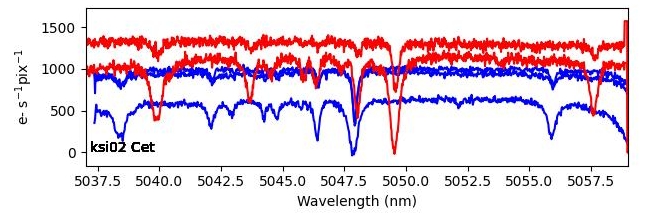}
 
      \caption{Comparison of the original CRIRES (blue) and CRIRES$^+$ (red) measured electrons per second per pixel as a function of wavelength  in the JHKLM bands. Multiple blue or red lines in the subpanels refer to distinct observations of the same target when the observing conditions were different. Spectra were not corrected for slit losses. The AO performance of CRIRES$^+$ is typcially the same or better than the original instrument. The CRIRES$^+$ data have been multiplied by 1.5 to account for the smaller pixel size for the new H2RG detectors (18 $\mu$m) to Aladdin InSb detectors pixels (27 $\mu$m) of the original CRIRES. Spectra were not corrected for either atmospheric telluric absorption or radial velocity shifts due to the Earth's motion. Wavelengths were taken from the FITS headers.
              }
         \label{fig:eff}
   \end{figure}
    The overall throughput of CRIRES$^+$ has been measured on 5 spectro-photometric standard stars (29 Psc, Ksi02 Cet, pi 02 Ori, eps Aqr and 42 Peg) by using the AO modes and the 0.4” slit width. The throughput was derived by scaling the observed spectrum, expressed in e-/s/pixel, by the theoretical  value estimated by multiplication of the known stellar brightness by the efficiency of the optical components. Fig.~\ref{fig:eff} shows an example of these measurements in the JHKLM-Bands comparing the throughput of the original CRIRES with the throughput of CRIRES$^+$. The observation of spectro-photometric standards is part of the instrument monitoring plan, and as such the results shown here should be regarded as preliminary. However, based on these preliminary results, the throughput is found to be about 10-15\% higher than the original CRIRES instrument. 
    
    Laboratory black body measurements of absolute throughput are subject to large errors, while on-sky, slit losses introduce large uncertainties. In addition there is considerable variation with band and with the shape of the grating blaze function for any setting within a band. Therefore, users interested in the absolute throughput for their observations should consult the CRIRES$^+$ User Manual and the CRIRES$^+$ Exposure Time Calculator that makes use of the best characterisation so far available.

\subsection{Spectral resolving power}
    The spectral resolving power $R$ for the 0.2” slit measured during the commissioning runs in 2021 was significantly lower than the expected resolving power of 100,000. Following an optimization of the camera focus in March 2022, the spectral resolving power of CRIRES$^+$ has been increased. The following values of the spectral resolving power were measured using observations of the OH sky emission lines with the 0.2” slit. For the H1582 wavelength setting: minimum $R$ = 100,000; average $R$ = 104,800 $\pm$ 2600 (echelle grating angle 66.3$^{\circ}$) and K2148 wavelength setting: minimum $R$ = 92,000; average $R$ = 96,500 $\pm$ 3200 (echelle grating angle 64.0$^{\circ}$). The spectral resolving power slightly increases across the science detectors in dispersion direction. There is some expected variation of the spectral resolving power with the echelle grating angle, and wavelength settings with a lower echelle grating angle value have a reduced spectral resolving power. The wavelength settings employing the lowest echelle grating angle (L3244, L3262,  M4187, M4211 and M4266) can exhibit minimum spectral resolving power values as low as $R$ = 86,000 (0.2” slit). All other wavelength settings exhibit minimum spectral resolving power values of at least $R$ = 92,000 (0.2” slit).  The effect of wavelength coverage and spectral resolving power on radial velocity studies is discussed in \cite{Bouchy2001A&A...374..733B}. Note that all values of $R$ that are specified here are for the 0.2” slit; when employing the 0.4” slit, these values need to be divided by two.  When the coherence time and AO correction are good the spectral resolution in some settings can exceed 100,000 as the PSF can be smaller than the minimum slit width of 0.2".

\subsection{Radial velocity (RV) precision}
\begin{figure*}
   \centering
   \includegraphics[width=18.5cm]{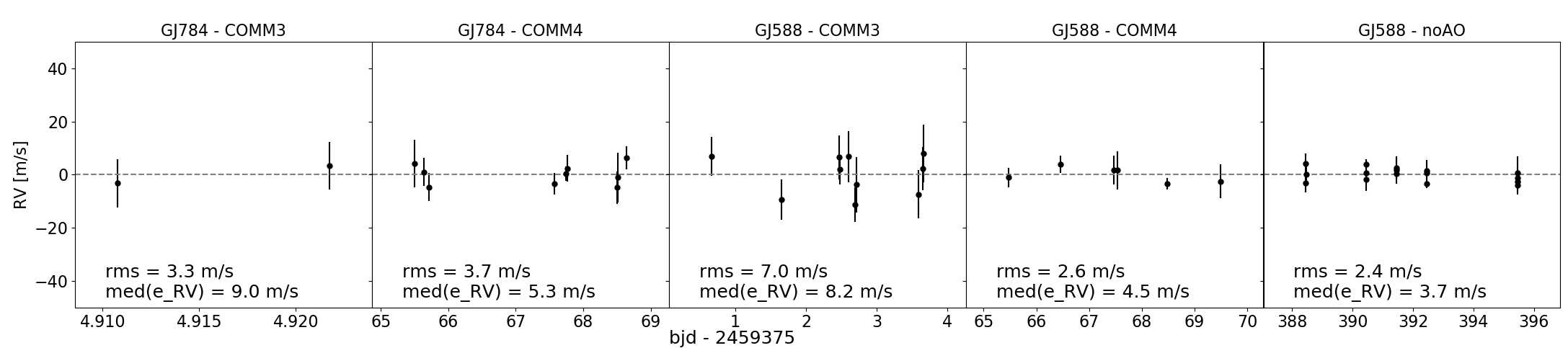}
      \caption{Radial velocities of GJ784 from June 2021 (first plot) and August 2021 (second plot) and of GJ588 from June 2021 (third plot) and August 2021 (fourth plot). The last plot shows the RV results of GJ588 for the Non-AO observations from July 2022. The standard deviation of the RV values varies from around 2.4~m~s$^{-1}$ to 7.0~m~s$^{-1}$. The error bars are based on the standard deviation of all used spectral orders of one observation and have a median value of around 5.2~m~s$^{-1}$.  For the COMM3 observations of GJ784 only two datapoints were taken.
              }
         \label{fig:RV_plot}
   \end{figure*}
   
\begin{figure}
   \centering
   \includegraphics[width=8.5 cm]{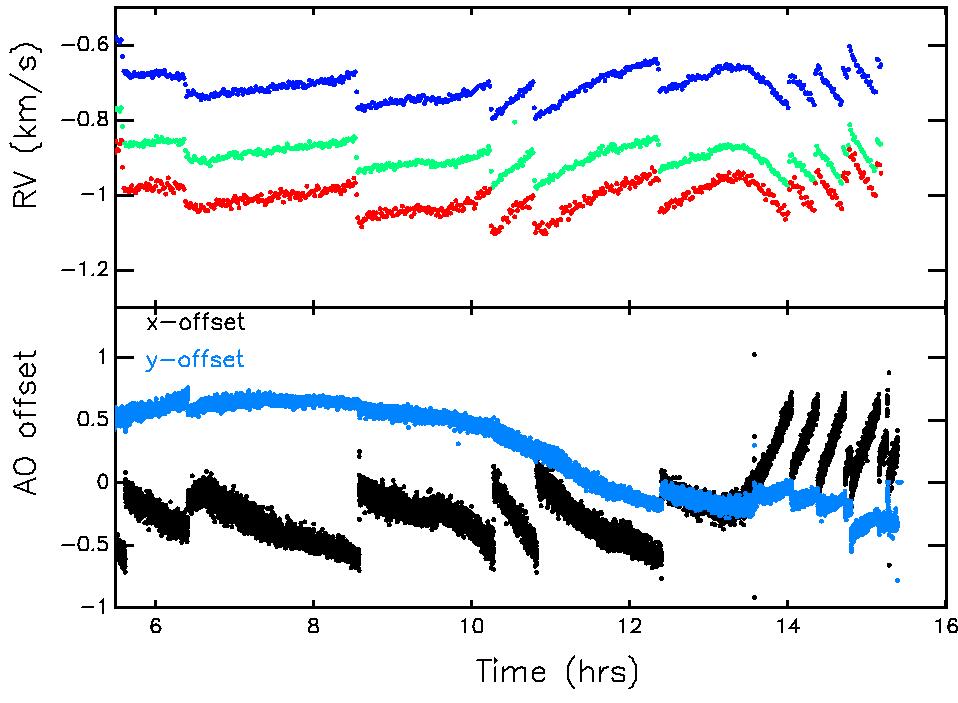}
      \caption{ Top panel: Radial velocity measured by cross correlation of gas cell features as a function of time for the lab observations in Garching using a fibre as an emission line source on top of which the gas cell was imprinted. Bottom panel: offsets applied to the adaptive optics system.}
         \label{fig:RV_plot_Garching}
   \end{figure}

The short gas cell (SGC) provides a stable long-term wavelength reference in the H and K bands. For a S/N of 150 per spectral pixel in the spectral continuum,  errors in  RV of 4~m~s$^{-1}$  over two hours have been attained for a sequence of short exposures (20 sec)  evenly spread over a 15 hour period in the laboratory by employing the short gas cell (SGC) as a simultaneous wavelength calibrator with the 0.2” slit ($R\approx100,000$).  The source was an illuminated fibre fed into the instrument. Adaptive optics was employed to keep the source on the slit. Jumps of several tens of m/s were found to correlate with movements of the AO system (see Fig. \ref{fig:RV_plot_Garching}). In operations such jumps have also been observed during the preset of the telescope although not during tracking.
    
On sky, the best RV precision that has been attained  using AO with CRIRES$^+$ is 2.6~m~s$^{-1}$ for GJ588 (K=4.76 mag) with DIT$\times$NDIT = 60~s and a S/N ratio of 280 in the brightest regions of the stellar continuum (see Fig.~\ref{fig:RV_plot}). In detail, for GJ588 there were 10 observations distributed over 4 nights during COMM3 (RMS=7.0 m~s$^{-1}$) and 6 observations over 5 nights in COMM4 (RMS=2.6 m~s$^{-1}$). In case of GJ784 there were 2 observations within one night during COMM3 (RMS=3.3 m~s$^{-1}$) and 9 observations over a time span of 4 nights during COMM4 (RMS=3.7 m~s$^{-1}$). All RV results in this paper were obtained using the {\sc viper} package \citep{Zechmeister2021} including the modeling of the telluric lines.

    The commissioning observations, as well as most of the other RV observations,  were done using the AO mode  as well as the gas cell and K2192 setting, while in July 2022 RV stability tests on GJ588 were performed  in NoAO mode with the gas cell. The results can be seen in the last plot of Fig.~\ref{fig:RV_plot}, showing a RV precision of around 2.4~m~s$^{-1}$ for 16 measurements taken over a time span of a week (DIT$\times$NDIT = 120~s). This result, plus the fact that the IP seems to be more symmetric and the instrument noise lower, raises the question of whether observations without AO mode could help to improve the RV precision.
    
    Due to slow drifts in the echelle gratings, the RV precision will degrade for longer exposure times. Errors of the RV measurements also depend on factors like the number of stellar absorption lines observed, the broadening due to the stellar rotation and the signal-to-noise ratio per pixel in the stellar continuum, and telluric line strength. Without a simultaneous wavelength calibrator  (SGC) during observations, the attainable RV precision will be much lower due to a slow drift of the Echelle grating. The drift in dispersion is of the order of 0.05 pixel over 30 min corresponding to an RV drift of about 50~m~s$^{-1}$.  Instead of taking attached wavelength calibrations with the UNe lamp and FPI, the telluric absorption lines of Earth’s atmosphere can also serve as a long-term stable, simultaneous wavelength reference. In most wavelength settings, a large number of these lines comes for free and will be directly imprinted onto the science data. \citet{2010A&A...515A.106F} demonstrated that telluric lines are intrinsically stable down to 10~m~s$^{-1}$ (rms). In any case, it is recommended to keep the exposure times short (DIT$\times$NDIT $\leq$ 120s) to minimize the effects of grating drifts. The thermal stability of the cryogenic instrument structure is $<$0.003 Kelvin and that of the echelle grating is $<$0.005 Kelvin also during movement of the functions. Both are equipped with an active control loop. We note that we have not found any direct correlation yet between these temperatures and the RV stability but the values are continuously monitored during operations. The radial velocity precision is still under investigation by the use of an RV standard taken $\sim$ weekly. The CRIRES$^+$ consortium will provide a publicly available radial velocity processing software for use with the CRIRES$^+$ instrument with the absorption cell, along with a user manual, test data sets and tutorials.

\subsection{Comparison of CRIRES$^{+}$ and UVES reduced spectra}

\begin{figure}
   \centering
   \includegraphics[width=9cm]{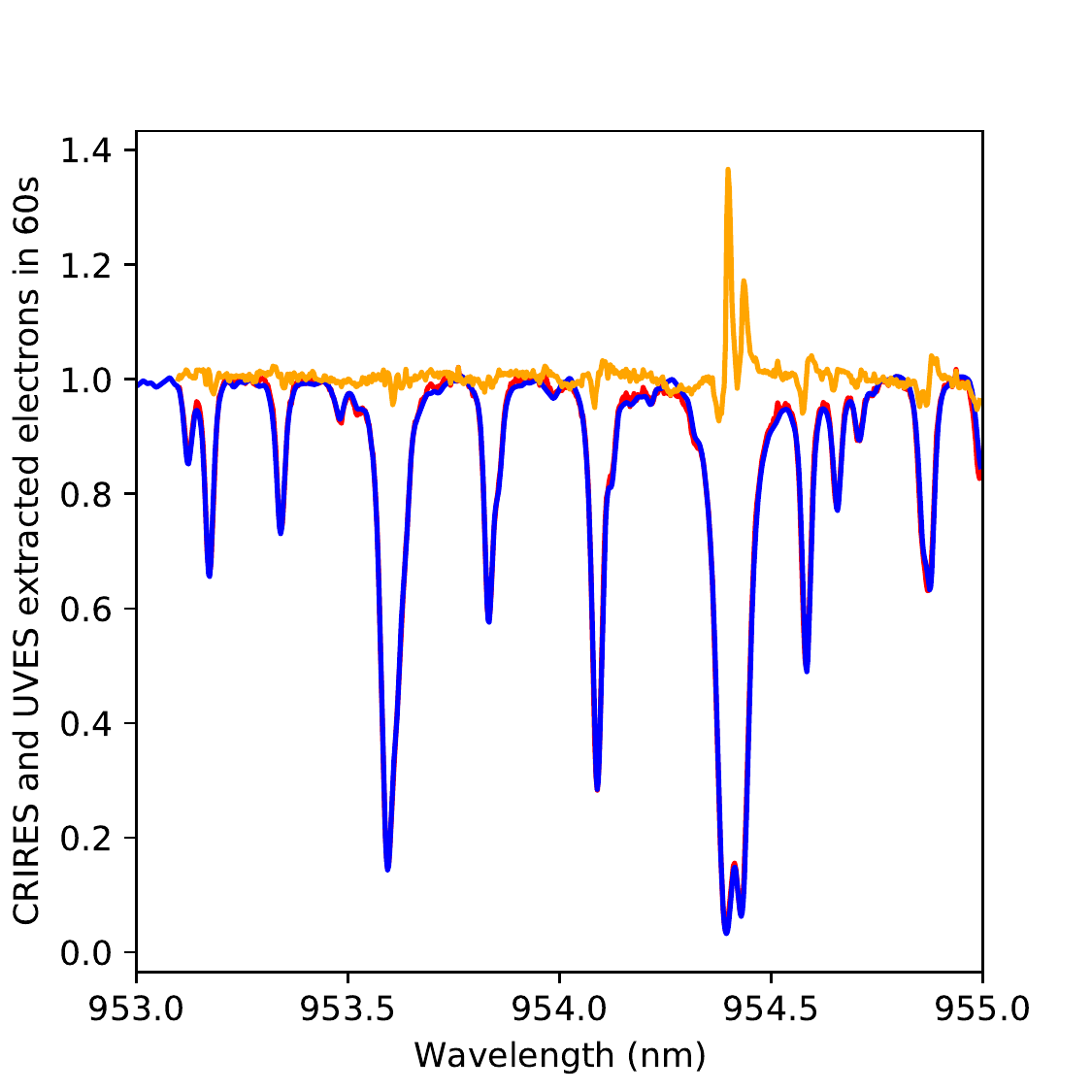}
      \caption{CRIRES$^{+}$  on UT3 (red) and UVES  on UT2 (blue) normalised reduced spectra towards HD\,33793, taken simultaneously with 60s exposure time using a 0.4$^{\prime\prime}$ slit. The UVES spectra have been smoothed to match the CRIRES$^{+}$ observations. The orange line shows the UVES spectrum divided by the CRIRES one. The CRIRES data were reduced using calibrations taken eight months after the observations with a 0.2$^{\prime\prime}$ slit and are shifted in wavelength to match the UVES observations. The UVES pipeline resamples the reduced spectra.
              }
         \label{fig:f_CRIRES_UVES}
   \end{figure}

   During instrument commissioning we observed the star HD\,33793 simultaneously with CRIRES$^+$  on UT2 in the Y1029 setting and UVES  on UT3 in the 860~nm setting (red arm only) for a sequence of five 60s exposures. Non-AO mode was employed for the CRIRES$^{+}$ observations with the slit width used for both instruments was 0.4$^{\prime\prime}$. The CRIRES$^{+}$ data were taken in stare mode as opposed to the normal nod-on-slit. The image quality of the observations derived from fitting a Gaussian to crosscuts in the cross dispersion direction was 0.66$^{\prime\prime}$ to 0.75$^{\prime\prime}$ for CRIRES$^{+}$ and 0.73$^{\prime\prime}$ to 0.78$^{\prime\prime}$ for UVES, not correcting for pixel size. The DIMM seeing (measured at $\lambda$=500~nm at the zenith) varied between 0.80$^{\prime\prime}$ and 0.87$^{\prime\prime}$ during the observations. Figure \ref{fig:f_CRIRES_UVES} shows a comparison between the normalized spectra derived from both spectrographs, with the UVES data being smoothed in order to match the CRIRES$^{+}$ spectral resolving power of $\sim$50,000. The comparison shows overall good agreement. These observations show that CRIRES$^{+}$ is more sensitive than UVES redwards of about 960~nm, compared with around 1000~nm with the original CRIRES instrument. 
   
\section{Conclusions}

    The ESO-led CRIRES$^+$ consortium consists of ESO (Germany, Chile), the Institut für Astrophysik, Georg-August Universität Göttingen (Germany), Thüringer Landessternwarte Tautenburg (Germany), Department for Physics and Astronomy, University of Uppsala (Sweden), INAF Osservatorio di Arcetri (Italy), INAF Osservatorio di Bologna (Italy). The instrument’s principal investigator is Artie Hatzes of the Thüringer Landessternwarte Tautenburg (Thuringian State Observatory Tautenburg), Germany. The co-investigators are Nikolai Piskunov (Uppsala University) and Ansgar Reiners (Georg-August Universität Göttingen). The formal CRIRES$^+$ Provisional Acceptance Europe took place on 20 September 2019. 

    From that time on the project experienced significant delays. Due to the unrest in Chile from October to December 2019, the originally planned CRIRES$^+$ installation was cancelled and shifted to January and February 2020. The first full system commissioning slots were initially scheduled for March and April 2020. However, the commissioning activities were put on hold due to the temporary closure of the Paranal Observatory due to COVID-19. With the ramping up of the Paranal Observatory in November 2020, the CRIRES$^+$ instrument was reactivated on 4 December 2020 and a series of remote tests was carried out. The upgraded CRIRES instrument was then commissioned during four runs. Due to the COVID-19 pandemic, the first two commissioning runs were executed remotely by employing the Garching Remote Access facility (G-RAF) to connect to the VLTI control room on Paranal. It was the first time that such a remote commissioning took place at ESO. 
    
    The call for proposals for the CRIRES$^+$ Science Verification (SV) projects was issued on 17 June and the SV run took place from 15 to 19 September 2021 \citep{2022Msngr.187...17L}. The observations were done from Paranal with support from Garching through G-RAF. The CRIRES$^+$ instrument has been in regular operations since 1 October 2021 (ESO Period 108) in dedicated modes. For ESO Period 109 starting in April 2022 the instrument was then offered in all modes. 

    The instrument largely fulfils the technical specifications, in some cases exceeding the goals. Performance has been shown to be at least as good as the original CRIRES. The scientific compliance of the instrument has been largely demonstrated, several aspects related to operations merit further investigation or require further development. Most of these issues do not lead to a violation of specifications but are obvious improvements that should be made in order to fulfil the instruments potential. All requested modes and standard wavelength settings have been commissioned. For spectral characterization of directly imaged planets, \citet{2021A&A...646A.150O} and \citet{2022arXiv220706436V} study the performance of a proposed HiRISE fibre coupling between the direct imager SPHERE \citep{2019A&A...631A.155B} and CRIRES$^+$. Using science verification data from CRIRES$^+$ \citet{2022AJ....164...79H} performed a atmospheric characterisation of an exoplanet - the ultra-hot Jupiter MASCARA-1b. In addition, \citet{2022arXiv220903669M} obtained transmission spectra of the Hot Saturn WASP-20b, observed in the K-band at a spectral resolving power of $\sim$92,000 during the first night of the Science Verification of the instrument which led to a tentative detection of H$_2$O. These findings demonstrate the scientific potential of CRIRES$^+$ and highlight the excellent opportunity for high-resolution atmospheric spectroscopy of diverse exoplanets. 

    The instrument is now fully integrated in Paranal operations. Technical time losses are very rare, and overheads are compliant with the specifications. The data archiving and access for CRIRES$^+$ is fully integrated into the ESO end-to-end operation scheme. All raw and reduced FITS data files are complaint with the ESO Data Interface Control Board (DICB). It is evident from the first period of operations that the DRS reduces each night and day’s data fast enough for operations. The data flow process has been optimized during the commissioning, science verification, and early operation and is now fully deployed at the observatory. In general, the instrument shows reliable operations and robust performance with no instrument hardware failures observed.

\begin{acknowledgements}
      The German contribution to the CRIRES$^+$ project was funded through the German Federal Ministry of Education and Research (Bundesministerium für Bildung und Forschung). The Swedish contribution was funded by a grant through the Knut and Alice Wallenberg Foundation. 
\end{acknowledgements}

\bibliographystyle{aa}
\bibliography{bibtex} 

\end{document}